\documentclass[reprint,aps,pre]{revtex4-1}

\usepackage{amssymb}
\usepackage{graphicx}
\usepackage{amsmath}
\usepackage{mathtools}
\usepackage{wasysym}
\usepackage{xcolor}
\usepackage{soul}    
\usepackage{epstopdf}
\usepackage[export]{adjustbox}
\usepackage{multirow,bigdelim,tablefootnote}
\usepackage{tcolorbox}

\newcommand{\be}{\begin{equation}}
\newcommand{\ee}{\end{equation}}
\newcommand{\ba}{\begin{eqnarray}}
\newcommand{\ea}{\end{eqnarray}}
\newcommand{\diff}[1]{\ensuremath{\operatorname{d}\!{#1}}}
\renewcommand{\epsilon}{\varepsilon}

\begin{document}

\title{
Structural changes in the Lennard-Jones supercooled liquid and ideal glass: 
an improved integral equation for the replica method}
\author{Jean-Marc Bomont}\email{jean-marc.bomont@univ-lorraine.fr}
\affiliation{Universit\'e de Lorraine, LCP-A2MC, UR\,3469, 1 Bd. Fran{\c c}ois Arago,
Metz F-57078, France}
\author{Dino Costa}
\affiliation{Dipartimento di Scienze Matematiche e Informatiche, Scienze Fisiche e Scienze della
Terra, Universit\`a degli Studi di Messina, Viale F. Stagno d’Alcontres 31, 98166 Messina,
Italy }
\author{Jean-Louis Bretonnet}
\affiliation{Universit\'e de Lorraine, LCP-A2MC, UR\,3469, 1 Bd. Fran{\c c}ois Arago,
Metz F-57078, France}
\author{Giorgio Pastore}
\affiliation{Universit\`a di Trieste, Dipartimento di Fisica, Strada Costiera 11, 34151
Grignano (Trieste), Italy}

\begin{abstract}

Framing the glass formation within standard statistical mechanics is an outstanding problem of condensed matter theory.
To provide new insight, we investigate the
structural properties of the Lennard-Jones fluid
in the very-low temperature regime,  by using a replicated version of the refined HMSA theory of the liquid state, combined with an appropriate split of the pair potential [Bomont \& Bretonnet, J.~Chem.~Phys. \textbf{114}, 4141 (2001)]. 
Our scheme 
allows one to reach an unprecedented low-temperature domain within both the supercooled liquid and the ideal-glass phase. 
Therein, a density-dependent temperature is identified, whereupon the radial distribution function experiences clear-cut structural changes, insofar as an additional peak 
develops in between the main and the second peaks. Such a structural feature points to a local structure of the Lennard-Jones ideal glass with an fcc-like short-range order, in the absence of any long-range order. 
     
\end{abstract}

\maketitle

\section{Introduction}\label{sec:1}
The ultimate fate of liquids at deep supercooling remains one of the key problems in classical physics and material sciences. Fundamental questions such as ``can any liquid be cooled down its freezing point to an isentropic temperature without vitrifying or crystallizing?''~\cite{Zanotto2018} or ``is there a low temperature limit for the existence of any supercooled liquid?'' stands at
the foundations of modern statistical physics of the liquid and glass states (see Ref.~\citenum{Zanotto2023} for a recent review). The glass transition~---~occurring at a temperature $T_{\rm g}$ below the freezing temperature~---~refers to the change from an equilibrium fluid state to a disordered solid state. Such amorphous solids, or glasses, are distinguished from crystalline solids by their lack of long-range structural order. 
It is usually believed that every liquid can be supercooled~---~avoiding crystallization~---~%
as long as a very fast cooling rate is applied. Good glass-formers are defined as those liquids for which crystallization can be avoided without resorting to exceedingly fast cooling rates~\cite{Kob1998}. 
Note that $T_{\rm g}$ is not uniquely defined and is not an equilibrium quantity, because it depends on the cooling rate~\cite{Guiselin2022}.

As a current standpoint, glasses are liquids that have left thermal equilibrium at $T_{\rm g}$ and, as such, they can no longer explore their space of configurations ergodically. On the microscopic scale, particle motion slows down so dramatically that it can be hardly detected, even on long timescales~\cite{Biroli2013}.
Since equilibration is hampered below $T_{\rm g}$, what happens down there still remains a matter of conjecture. If one considers that such a nearly-arrested system behaves macroscopically like a solid, despite its microscopic structure still resembling that of a completely disordered liquid, it is not surprising that very different theoretical interpretations were suggested over time, based on extensions of liquid state theory, and providing equally good explanations of available data
(see Refs.~\citenum{Angell1995,Debenedetti2001,BerthierBiroli2011} for extensive reviews). 
In this context, a first prominent approach is the mode-coupling theory (MCT)~\cite{Gotze2009}, that is able to predict the drastic increase of the relaxation time upon supercooling, solely from structural information. Approximately mimicking the dynamics of supercooled liquids, where mobile regions (in which the particles relax on a microscopic timescale) coexist with immobile regions (where relaxation is very slow), MCT predicts that the structural relaxation time scales as a power law with temperature, ultimately diverging at an ideal-glass transition~\cite{Royall}. 
In this framework, most of studied models, such as the well-known Kob-Andersen model, require the introduction of a certain degree of polydispersity to avoid crystallization upon cooling~\cite{Kob1994,Kob1995A,Kob1995B}.

On the other hand, the idea that a glass state of matter could exist under equilibrium conditions is a long-standing concept, dating back to the 80-year-old Kauzmann's seminal article~\cite{Kauzmann1948}. 
The hypothesis that an equilibrium phase transition at $T_{\rm K}$ is underlying the formation of glasses is still a major current challenge but remains a matter 
of debate~\cite{Stillinger,Tanaka2003,BerthierBiroli2011,Zanotto2023,Martin2025}.

A transition at a finite, non-zero, $T_{\rm K}$ towards a hypothetical ``ideal glass'' remains to be demonstrated by direct measurements. From the computational point of view, such an objective is thought to require the development of novel and substantially more sophisticated numerical algorithms than those currently available to achieve equilibration closer to the putative Kauzmann temperature~\cite{Guiselin2022,Royall2018}.

Recent state-of-the-art computer studies and theoretical analysis have provided support to the hypothesis that supercooled liquids, carefully maintained in equilibrium conditions at low temperatures, can undergo a discontinuous liquid-glass phase transition, questioning in this way the assumption that the glass formation is essentially a dynamic process accompanied by small structural changes~\cite{Guiselin2022}. 
 
In this context, a fundamentally thermodynamic perspective of the glass transition was provided by the so-called ``random first-order transition'' (RFOT) theory~\cite{Wolynes}, that includes the elegant ``replica method'', which purpose is to locate 
the ideal-glass transition in the density-temperature plane~\cite{Franz1995,Mezard1996,Franz1997,Franz1998,Cardenas1998,Cardenas1999,Coluzzi1999}. Such a static approach involves Integral Equation Theories (IET)~\cite{BomontACP2008,Hansen2013,Janssen2024}, and gave rise, since then, to many studies
(see Ref.~\citenum{Berthier2022} for an extensive review). The replica method requires the use of an appropriate order parameter $Q$, which quantifies the overlap between two coupled clones of a system and therefore can distinguish between the equilibrium liquid and the glass phases. The overlap is also useful above the glass transition, since it has been suggested that the RFOT might be preceded by a ``precursor transition'' occurring at higher temperatures, signaling the coexistence of two phases corresponding to low and high values of $Q$. The low-$Q$ phase is associated with the supercooled liquid, while the high-$Q$ phase is for the glass. Later on, a simplified but equivalent scheme was proposed by two of us in collaboration with Jean-Pierre Hansen~\cite{BomontEPL2014} and, since then, successfully applied to a variety of interaction models~\cite{BomontJCP2014,BomontMOLPHYS2015,BomontPRE2015,%
Bomont2017,BomontJCP2017,BomontJCP2019,BomontPRE2022,BomontJCP2024}. 

In particular, we compared the predictions coming from our simplified scheme~\cite{BomontEPL2014} with those obtained within the original theory~\cite{Franz2013} for two different interaction models, namely soft spheres~\cite{BomontJCP2019} and Lennard-Jones~\cite{BomontJCP2024}. Thereby, we showed that, under strictly identical conditions, the two approaches provide identical results for both models, as for the dynamical transition.

In the present work, we assume that thermal equilibrium can be maintained at an arbitrary low temperature, and we study how deeply a liquid can be supercooled within the framework of IET. In particular, we aim to investigate the fate of structural correlations within the deeply supercooled liquid and the ideal-glass phase.
To our purposes, we study the Lennard-Jones model by a replicated version of the thermodynamically self-consistent hybridized-mean-spherical approximation theory (R-HMSA)~\cite{Zerah}, combined with the Optimized Division Scheme for the pair interaction proposed by two of us~\cite{Bomont2001}, that is known to provide accurate predictions for the structure and thermodynamics of simple liquids~\cite{Bomont1997,Bomont1998}. At a first stage, upon cooling the system along several isochores, thermodynamically self-consistent results are found in ranges of low temperatures never reached before by any other computational tool, either in the liquid or in the ideal-glass phases. Within this regime, we identify the upper boundary, given by the dynamical transition temperature $T_{\rm D}$, for the existence of the 
ideal-glass phase. Then, by free energy calculations, we locate the critical temperature $T_{\rm cr}<T_{\rm D}$ below which the ideal-glass phase is the stable one. Finally, this offers us the opportunity to identify in both phases a density-dependent low temperature threshold $T_0<T_{\rm cr}$ across which a clear-cut structural change occurs.
 
Our choice for the pure Lennard-Jones, in comparison for example with a standard glass former such as the Kob-Andersen mixture~\cite{Kob1994,Kob1995A,Kob1995B} is dictated not only by the possibility to characterize the glass behavior of perhaps the most exhaustively studied model for the fluid state. In fact, since in our formalism the study of a pure fluid involves the introduction of a symmetric binary mixture, starting from a binary mixture at the outset, would drastically increase the complexity of the scheme, requiring in particular several different relationships to let us predict the pair structure of the fluid in a thermodynamically consistent manner.  

The paper is organized as follows: 
in Section~\ref{sec:2} we review the essentials of our IET approach and RFOT scenario. 
In Section~\ref{sec:3} the interplay between supercooled liquid and ideal glass is investigated. 
In Section~\ref{sec:4}, we present our results for the structure within the liquid and ideal-glass phases. Our conclusions follow in Section~\ref{sec:conclusions}.

\section{THEORETICAL BACKGROUND}\label{sec:2}
\subsection{Pair interactions}
We consider a symmetric binary ``mixture'' of two weakly coupled clones (or replicas, labeled 1 and 2), each consisting of $N$ atoms interacting via pair potentials $u_{11}(r)\equiv u_{22}(r)=u(r)$, while atoms of different clones interact via a weak attractive potential $u_{12}(r)$. The total potential energy hence reads :
\begin{eqnarray}
V_{_{N,N}}& = & \sum_{i} \sum_{j>i} u_{11}\left( \left\vert \mathbf{r}%
_{i}^{1}-\mathbf{r}_{j}^{1}\right\vert \right) +\sum_{i} \sum_{j>i}
u_{22}\left( \left\vert \mathbf{r}_{i}^{2}-\mathbf{r}_{j}^{2}\right\vert
\right)  \notag \\
& \quad & +\sum_{i} \sum_{j} u_{12}\left( \left\vert \mathbf{r}_{i}^{1}-%
\mathbf{r}_{j}^{2}\right\vert \right)  \label{EQ1}
\end{eqnarray}

The intra-replica interactions are given by the Lennard-Jones (LJ) potential:
\be
u(r)=4\varepsilon \left[ \left( \dfrac{\sigma }{r}\right) ^{12}-\left( \dfrac{%
\sigma }{r}\right) ^{6}\right]\,,  \label{eq:LJ}
\ee
where the effective diameter $\sigma$ is the distance at which $u(r)=0$, and $\epsilon$ is the potential 
well depth at $r_{\rm min}=\sigma\sqrt[6]{2}$ (see Fig.~\ref{fig:PotSplit}). Using $\sigma$ and $\epsilon$ as units of length and energy, respectively, the reduced temperature is defined as $T^\ast=k_{\rm B}T/\epsilon$ (with $k_{\rm B}$ the Boltzmann constant) and the reduced density $\rho^\ast=\rho\sigma^3$; finally the inverse temperature is $\beta=1/T^\ast$. Atoms belonging to different replicas are weakly coupled through a short-range inter-replica attraction in the form~\cite{Mezard2012}: 

\begin{equation}
u_{12}(r)=\varepsilon _{12}w(r)=-\varepsilon _{12}\left[ \frac{(c/\sigma)^{2}}{%
(r/\sigma )^{2}+(c/\sigma)^{2}}\right]^{6},  \label{eq:u12}
\end{equation}%
where $c$ is chosen such that the range of the attraction is significantly smaller than the mean distance between neighbouring atoms,  $d/\sigma \approx \sqrt[3]{1/\rho^{\ast}}$ (see inset of Fig. 1), to ensure that, due to the strong repulsion between
atoms of the same replica, an atom of one replica can interact with at most one atom of the other replica. The attraction induces pairing of atoms of opposite clones into diatomic ``molecules'', so that the symmetric binary mixture looks like a ``molecular liquid''. Notice that the exact form of the chosen $u_{12}(r)$ is irrelevant since we shall be eventually interested in the limit $\epsilon_{12}\to 0$.

\subsection{HMSA scheme and ODS splitting procedure}
For a given state point, the equilibrium pair structure of the symmetric binary
mixture is characterized by two pair distribution functions $g_{11}(r)\equiv
g_{22}(r)$\ and $g_{12}(r)$\ which can be calculated approximately by
solving the two coupled integral equations resulting from the two Ornstein-Zernike\
relations between the $g_{ij}(r)$\ and the direct and indirect correlation functions $%
c_{ij}(r)$, $\gamma_{ij}(r)=g_{ij}(r)-1-c_{ij}(r)$ respectively, upon use of approximate closure relations~\cite{Hansen2013}.

To our purposes, we choose the HMSA closure~\cite{Zerah}, whose radial distribution function for the symmetric binary-mixture reads
\begin{figure}[t]
\begin{tabular}{c}
\includegraphics[width=0.48\textwidth]{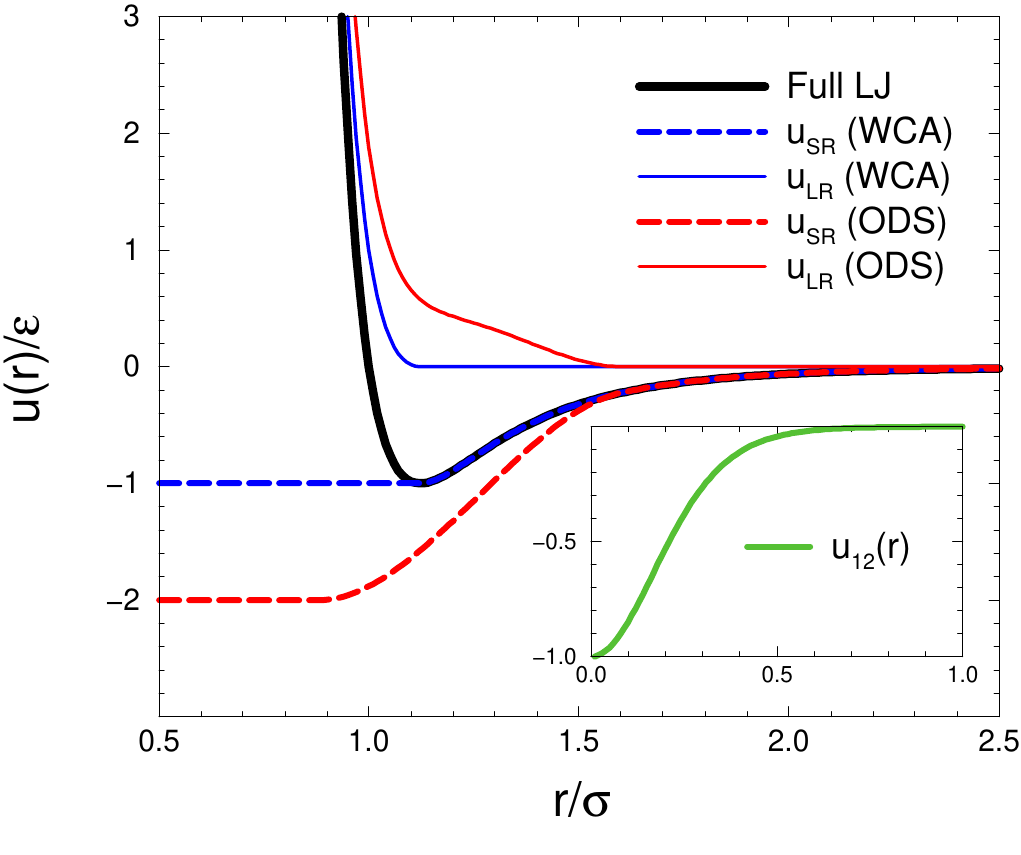}  
\end{tabular}%
\caption{LJ potential in units of $\varepsilon$ (black), as split according to either the WCA (blue)
or ODS (red) schemes. Inset: $u_{12}(r)$ in the same units, as drawn with $c=0.6\sigma$ and $\epsilon_{12}=\epsilon$.}
\label{fig:PotSplit}
\end{figure}
\begin{eqnarray}
g_{ij}(r)&=& \exp \left[ -\beta u_{ij}^{\rm SR}(r)\right]  \notag \\
&\quad&\times \left[ 1+\frac{\exp \left\{ f_{ij}(r)\left[
\gamma_{ij}^\ast(r)\right] \right\} -1}{f_{ij}(r)}\right]. \label{eq:HMSA}
\end{eqnarray}
Equation~(\ref{eq:HMSA}) requires \textit{i}) a suitable separation of the pair potentials into a reference short-range and a perturbative long-range part, so that $u_{ij}(r)=u_{ij}^{\rm SR}(r)+u_{ij}^{\rm LR}(r)$ and \textit{ii}) the use of the "renormalized" indirect correlation function $\gamma_{ij}^\ast(r)=\gamma_{ij}(r)-\beta u^{\rm LR}(r)$~\cite{Lee1996} as involved in other closures~\cite{DH1,DH2,Restrepo1,Restrepo2,Vompe1,Vompe2}.
The ``switching functions'' $f_{ij}(r)=1-\exp(-\alpha _{ij}r) $ depend on the inverse range parameters $\alpha _{11}=\alpha _{22}$\ and $\alpha _{12}$. As based on our previous experience~\cite{BomontEPL2014}, we have followed here the most convenient choice $\alpha_{12}\rightarrow \infty$, i.e., $f_{12}(r)=1$.
We recall that in the opposite limits $\alpha_{ij}\rightarrow 0$ [i.e., $f_{ij}(r)=0$], or $\alpha_{ij}\rightarrow \infty$ [i.e., $f_{ij}(r)=1$], the soft-core mean-spherical approximation (SMSA) and the thermodynamically inconsistent HNC are respectively recovered. In our scheme, we use local consistency so that the state-dependent parameter $\alpha_{11}$ is adjusted to ensure thermodynamic self-consistency between virial and long-wavelength compressibility equations of state.

Although in the HMSA the splitting procedure of the pair potential into short-range and long-range contributions, is quite arbitrary, it turns out to sensitively influence the quality of theoretical predictions. 
Originally, the mostly used Weeks-Chandler-Andersen prescription (WCA)~\cite{WCA}, for the LJ potential was followed, see  Fig.~\ref{fig:PotSplit}. Rather surprisingly, when the WCA split is used, HMSA provides bridge and cavity functions in the core region (especially at zero separation) not in good agreement with MC data, over a wide range of densities and temperatures~\cite{Bomont1998}. 
Nevertheless, such a requirement turns to be crucial to get reliable predictions for thermodynamic properties, such as the chemical potential~\cite{Bomont_2003_JCP_1,Bomont_2003_molphys,Bomont_2003_JCP_2,Bomont_2004_JCP,Bomont_2006_JCP,Bomont_2007_JCP,Lee_c_2010}, pointing out how the thermodynamic consistency alone is not a sufficient requirement to produce accurate theoretical predictions. 

In this concern, two of us proposed a different splitting procedure for LJ, 
the so-called Optimized Division Scheme (ODS)~\cite{Bomont2001}.
The HMSA/ODS scheme was shown to compare well with Monte Carlo data,
as for structural and thermodynamic properties of the LJ fluid~\cite{Bomont2001,Roland2024}. The special feature pointed out in that study was the extreme sensitivity of the bridge function to the perturbative part of the pair potential in the core region. 
In the ODS, the LJ 
$u^{\rm LR}(r)$ is constant for small $r$, as in the WCA scheme, and then joins smoothly $u(r)$ for larger $r$, by means of a fourth order polynomial (see again Fig~\ref{fig:PotSplit}):   
\be
u^{\rm LR}(r)=
\begin{cases*}
-p\epsilon & if $r\leq r_1$ \\
a_1+a_2 r+a_3 r^2+a_4 r^3 & if $r_1 < r \leq r_2$\\
u(r) & otherwise\,,
\end{cases*}
\ee
%
in which $r_{2}=r_{\rm min}+2( r_{\rm min}-r_{1})$, with $r_{\rm min}=\sigma \sqrt[6]{2}$, 
$r_{1}=0.88\sigma$ and $r_{2}=1.6\sigma$.
An important difference with respect to the WCA split, where $u^{\rm LR}(r=0)=-\varepsilon$, is that in ODS $u^{\rm LR}(0)=-p\epsilon$. Then, the choice $p=2$ was motivated by the fact that the value of the bridge function
at zero distance is connected~---~via the zero-separation theorem~\cite{Lee1996}~---~to the insertion of a single particle with twice the strength $\varepsilon$ of the interaction. ODS reduces to WCA for $p=1$ and $r_{1}=r_{2}=r_{\rm min}$.
The parameters $a_{1}$ to $a_{4}$
are determined such that $u_{\rm LR}(r)$ is continuous and has continuous first derivative $u^{\prime
}_{\rm LR}(r)$ both at $r_{1}$ and $r_{2}$;
because of typos in Refs.~\citenum{BomontACP2008} and~\citenum{Bomont2001}, we prefer to rewrite here their correct expressions:
\begin{widetext}
\begin{subequations}
\begin{align}
a_{1}&=\dfrac{r_{1}^{3}u(r_{2})-r_{2}r_{1}^{3}u^{\prime
}(r_{2})-3r_{2}r_{1}^{2}u(r_{2})+r_{1}^{2}r_{2}^{2}u^{\prime
}(r_{2})+3p\varepsilon r_{1}r_{2}^{2}-p\varepsilon r_{2}^{3}}{%
(r_{1}-r_{2})^{3}}\,, \\ 
a_{2}&=-\dfrac{r_{1}[-r_{1}^{2}u^{\prime }(r_{2})-r_{1}r_{2}u^{\prime
}(r_{2})+6p\varepsilon r_{2}-6r_{2}u(r_{2})+2r_{2}^{2}u^{\prime }(r_{2})]}{%
(r_{1}-r_{2})^{3}}\,, \\ 
a_{3}&=\dfrac{-2r_{1}^{2}u^{\prime }(r_{2})+3p\varepsilon
r_{1}-3r_{1}u(r_{2})+r_{2}r_{1}u^{\prime }(r_{2})+3p\varepsilon
r_{2}-3r_{2}u(r_{2})+r_{2}^{2}u^{\prime }(r_{2})}{(r_{1}-r_{2})^{3}}\,, \\ 
a_4 &=-\dfrac{-r_{1}u^{\prime }(r_{2})-2u(r_{2})+r_{2}u^{\prime
}(r_{2})+2p\varepsilon }{(r_{1}-r_{2})^{3}}\,,%
\end{align}
\end{subequations}
\end{widetext}

As for $u_{12}(r)$ of Eq.~(\ref{eq:u12}), its short-range nature, together with our position 
$f_{12}(r)=1$ in Eq.~(\ref{eq:HMSA}), simply implies $u^{\rm SR}_{12}(r) \equiv u_{12}(r)$ and therefore 
$u^{\rm LR}_{12}(r)=0$.

Our simplified replica method involves, for given temperature and density, the solution of the coupled set of equations given by the Ornstein-Zernike relations for the symmetric binary mixture, together with the HMSA
closures of Eq.~(\ref{eq:HMSA}), within the ODS split of Eq.~(5) for the LJ potential.
For clarity, such a scheme will be called in the following R-HMSA (replicated HMSA). It provides "non-trivial" solutions, namely $g_{11}(r)$ and $g_{12}(r)$, exactly as in Eq.~(\ref{eq:HMSA}). These predictions will be constantly compared to the "trivial" HMSA $g(r)$ of the one-component system in the same thermodynamic conditions, and within the same ODS split.

We have solved numerically the equations for the pair-structure by the efficient Gillan iterative algorithm~\cite{Gillan1979}. We have verified the robustness of our predictions against several resolutions $\Delta r$ of the spatial grid, and corresponding total number $N_{\rm g}$ of grid points. In practice, to keep the overall range in direct and reciprocal space large enough (and the grid spacing small enough) and to minimize truncation and discretization errors. In practice, $N_{\rm g}$ is never less than $4097$ and $\Delta r$ is never larger than $0.01\sigma$.

\subsection{RFOT scenario}\label{subsec:RFOT}

In an ideal glass, atoms vibrate around disordered equilibrium positions, rather than around the periodic positions of a crystal lattice. In this disordered state, the system is ``stuck'' in the lowest minimum of the suitably defined free energy  landscape~\cite{Charbonneau2014,Bretonnet2016}. Although a detailed probe of such a landscape~---~characterized by an exponentially large number of local minima~---~is clearly unachievable, the replica method for the study of disordered systems~\cite{Mezard1987} can be put to good use to probe such a landscape, 
by defining a suitable order parameter~\cite{Franz1998,Monasson1995} that can be calculated, at least approximately, using techniques borrowed from the theory of liquids~\cite{Hansen2013}. 
Specifically, 
the inter-replica $g_{12}(r)$ of Eq.~(\ref{eq:HMSA}) allows the calculation of the overlap $Q$,
\begin{equation}
Q=8\pi \rho  \overset{\infty }{%
\underset{0}{\int }}g_{12}(r)w(r)r^{2} \diff r\,,  \label{eq:Q}
\end{equation}%
with both $Q$ and $g_{12}(r=0)$ as order parameters providing the key diagnostics tools to locate the RFOT.

\begin{figure*}[t!]
\begin{tabular}{ccc}
\includegraphics[width=0.45\textwidth]{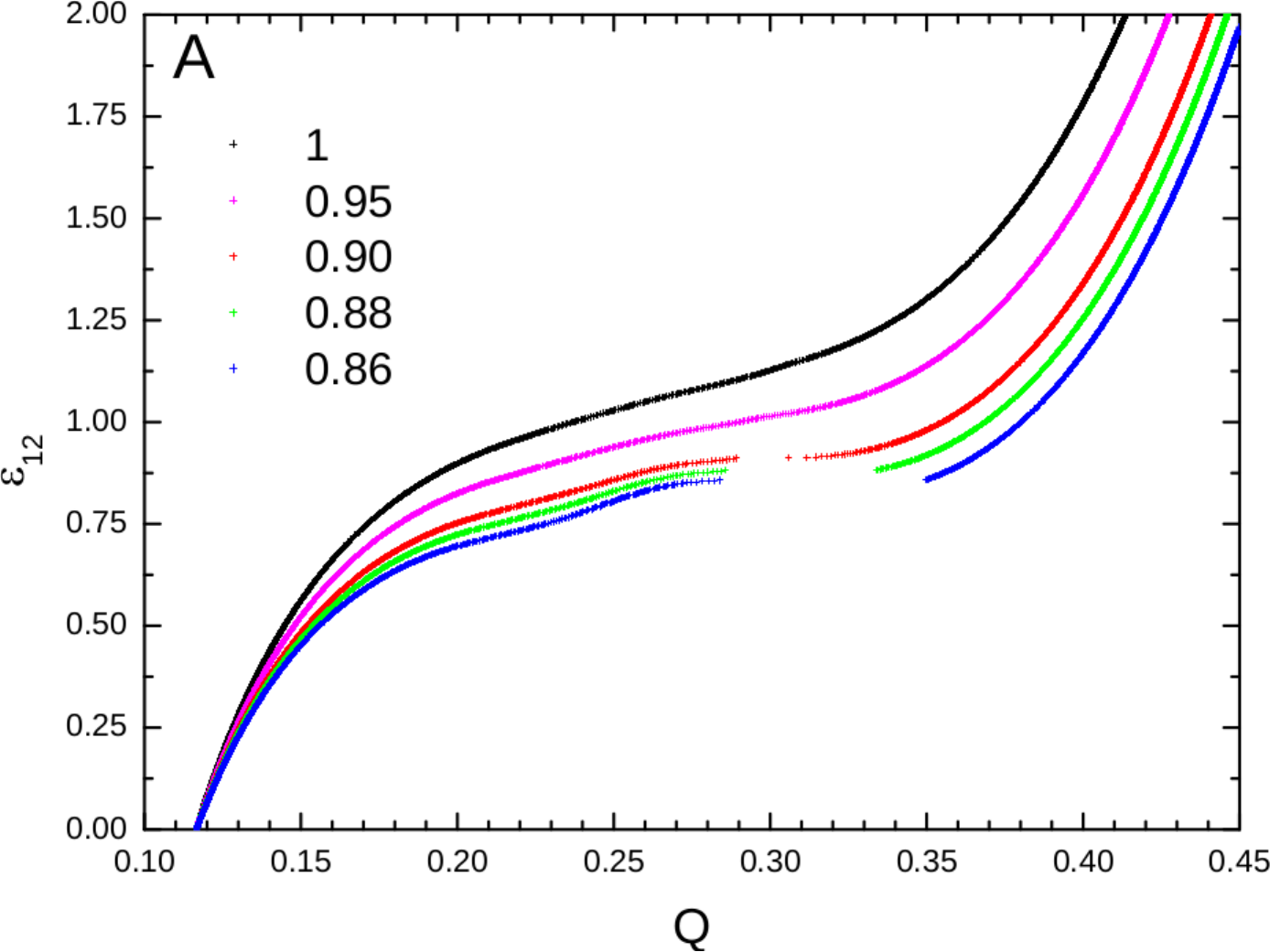} & 
\includegraphics[width=0.43\textwidth]{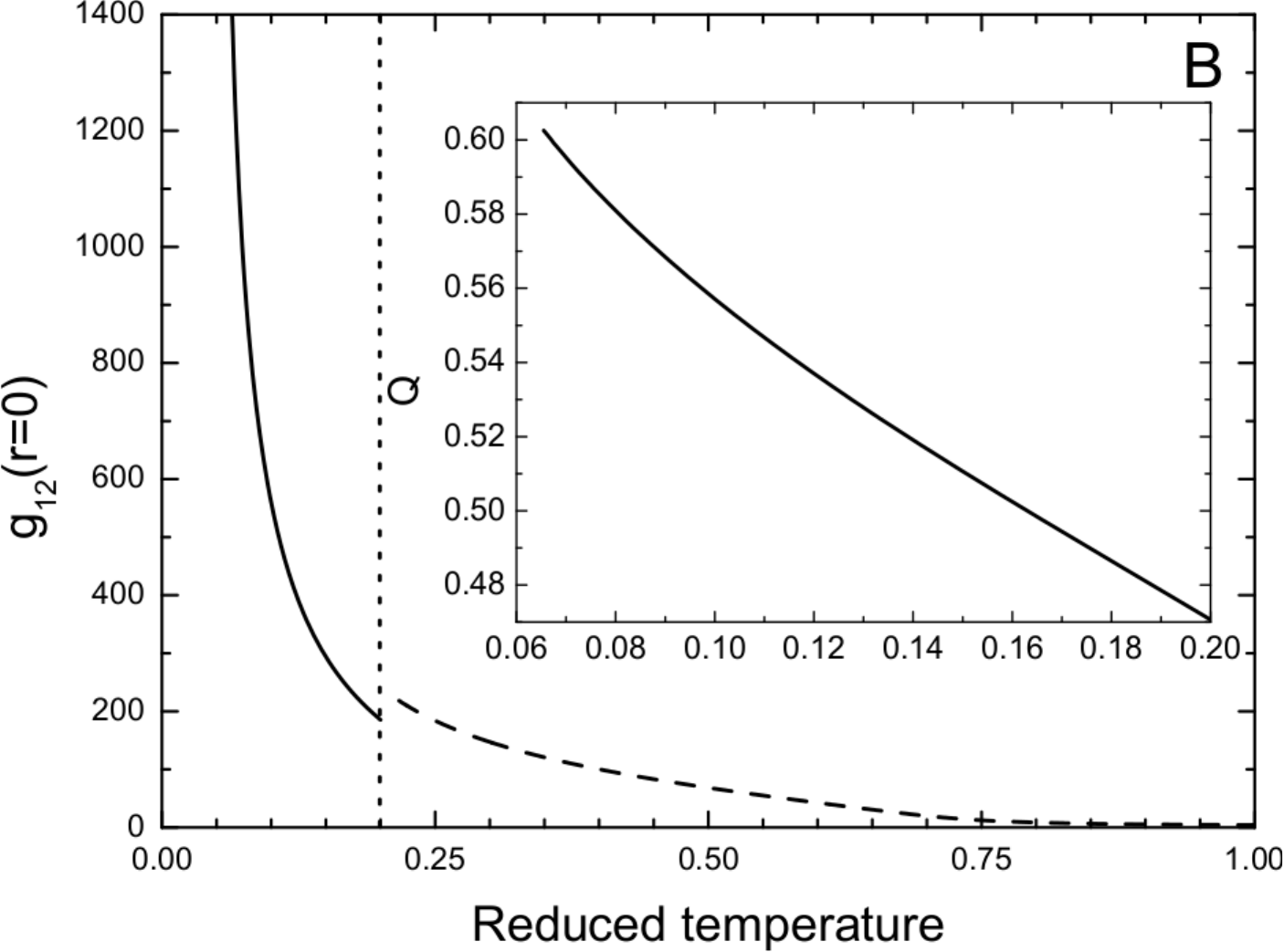} 
\end{tabular}
\caption{A: dependence of the order parameter $Q$ on the inter-replicas coupling
$\epsilon_{12}$, along several isotherms (in the legend) for $\rho^\ast=1$. A discontinuous jump of $Q$ is observed at $T^\ast=0.88$, indicating the occurrence of the expected ``precursor transition''. In this representation, the behavior is reminiscent of isotherms in a liquid-vapor system. B: behavior of the order parameter $g_{\mathrm{12}}(r=0)$ for $\rho^\ast=0.9$ upon cooling, under condition $\epsilon_{12}=\epsilon$ (dashed line) and after extinction of $\epsilon_{12}$ at $T^\ast=0.2$ (full line). The inset shows the overlap $Q$ for $T^\ast<0.2$.} 

\label{fig:protocol}
\end{figure*}

Our search protocol consists in solving the R-HMSA for the symmetric binary mixture at a given temperature $T^\ast$, starting from an initial finite value of the inter-replica coupling $\epsilon_{12}$ of Eq.~(\ref{eq:u12}). Then, the evolution of $Q$ and $g_{12}(0)$ is followed upon cooling.
Along a given isochore, at reasonably high temperature, 
the system is expected to be in its liquid phase. Therefore,
once $\epsilon_{12}$ is switched off,
the two replicas completely decouple, as signaled by
$Q$ taking its random value $Q_{\rm random}$ and $g_{12}(0)=1$.
In this condition, perfectly equivalent predictions are obtained by using either R-HMSA (for the symmetric binary mixture) or HMSA (for the one-component system).
On the other hand, $Q$ and $g_{12}(0)$ increase upon cooling and when $\epsilon_{12}$ is switched off at a sufficiently low temperature, $Q \gg Q_{\rm random}$ and $g_{12}(0)\gg1$. This in turn
implies that the two replicas remain correlated, even in the absence of coupling. In other words, the system falls in 
its ideal-glass phase, where replicas are all trapped within the same minimum of the free energy landscape. In this condition, R-HMSA and HMSA predictions,
as concerning different states attained by the system, in principle differ.
The highest temperature at which R-HMSA "non-trivial" solutions survive, identifies the so-called dynamical transition temperature $T^\ast_{\rm D}$.

The threshold $T^\ast_{\rm D}$ sets the upper limit for the hypothetical existence of the ideal glass, above which
the system falls in its liquid phase. Below, however, $T^\ast_{\rm D}$ says nothing
about the relative stability of glass and liquid,
to be formally discriminated by comparing the excess free energies $F_{\rm ex}$ along both branches. 
Then, the stable phase for given thermodynamic conditions, is identified as the one having the lower 
free-energy content.
As for the supercooled liquid, we have calculated $F_{\rm ex}^{\rm L}$ by means of a standard two-path thermodynamic integration scheme (see Ref.~\citenum{Costa2002} for details). In the first step, starting from low density $(\rho^\ast=0.05)$, virial pressure predictions are collected up to $\rho^\ast=1.0$, along the supercritical isotherm $T^\ast=4.0$, with an increment $\Delta\rho^\ast=0.005$. In the second step, internal energies are recorded along several isochores, by decreasing the temperature in steps $\Delta T^\ast=0.005$. 
A three-step protocol is instead applied to determine the free energy of the ideal glass, 
$F_{\rm ex}^{\rm G}$: 
the first step is carried out for a fixed value of $\epsilon_{12}$ within the "molecular liquid" along the isotherm $T^\ast=4.0$ up to $\rho^\ast=1.0$. In the second step, the ``molecular liquid'' is cooled down along several isochores and at a certain temperature $\epsilon_{12}$ is progressively reduced to zero. In the last step, the ideal-glass branch is mapped out by keeping $\epsilon_{12}=0$, while lowering the temperature. This protocol, with $c=0.6\sigma$, provides a continuous thermodynamic path over which $F_{\rm ex}^{\rm G}$ can be calculated.

\section{Interplay between supercooled liquid and ideal glass}\label{sec:3}

A key prediction of the RFOT theory is that, at a given density, a precursor of the transition can be already traced back in the behavior of $\varepsilon _{12}$ vs $Q$ 
in the liquid phase~\cite{BerthierPRE2015}. In fact, while $Q$ increases smoothly with $\varepsilon_{12}$ at high temperature, it is expected to develop a sharp jump for lower $T^\ast$, such a feature heralding the incipient RFOT itself. This peculiar behavior is shown in Fig.~\ref{fig:protocol}A. Therein, the variation of $Q$ with $\epsilon_{12}$ is monitored, using a very small increment $\Delta \epsilon_{12}=0.001$, along several isotherms in the range $0.86 <T^\ast<1.0$, with $\rho^\ast=1.0$ and $c = 0.6\sigma$ in the inter-replica potential of Eq.~(\ref{eq:u12}). For $\epsilon_{12}=0$, the random overlap value $Q_{\rm random}= 0.1166$ is invariably recovered for all $T^\ast$. Then, $Q$ tends to increase less and less rapidly with $\epsilon_{12}$ as $T^\ast$ decreases. At $T^\ast=0.88$, a clear-cut discontinuity occurs at $\epsilon_{12}= 0.88$, where $Q$ jumps from $0.285$ to $0.334$, before increasing further. The discontinuity $\Delta Q$ increases upon cooling. 
In the figure, the reverse behavior of $\epsilon_{12}$ vs $Q$ is shown, to conveniently highlight the analogy with the standard liquid-gas coexistence region: in the present case, low-$Q$ and high-$Q$ states coexist below the ``critical point'' of the precursor transition, marked by a zero-slope inflection point of the $\epsilon_{12}$ vs $Q$ curve; the corresponding critical temperature lies between $T^\ast= 0.88$ and $T^\ast= 0.90$.

To investigate the ideal-glass phase, we have solved the R-HMSA for $g_{\mathrm{11}}(r)$ and $g_{\mathrm{12}}(r)$ upon cooling, along several isochores in the range $0.85 \leq \rho^\ast \leq 1$. In details, we have used for each density two initial values for the inter-replica coupling at high temperature, namely $\epsilon_{12}=0$ and $\epsilon_{12}=\epsilon$. 
Upon cooling the mixture with $\epsilon_{12}=0$, since replicas are uncorrelated, $g_{\mathrm{12}}(0)=1$ and $Q =Q_{\rm random}$. In this case, as we said, R-HMSA and HMSA provide perfectly equivalent predictions. As for $\epsilon_{12}=\epsilon$, the evolution of $g_{\mathrm{12}}(0)$ with $T^\ast$ and fixed $\rho^\ast=0.90$ is shown by the dashed line in Fig.~\ref{fig:protocol}B: we see that $g_{\mathrm{12}}(0)$ increases with decreasing $T^\ast$,
thereby signaling that replicas progressively correlate upon cooling. Then, $\epsilon_{12}$ is progressively switched off at $T^\ast=0.2$ (signaled by the vertical dotted line in Fig.~\ref{fig:protocol}B), where indeed $g_{\mathrm{12}}(0) \gg 1$ and $Q \gg Q_{\rm random}$. 

\begin{figure}[t]
\begin{tabular}{c}
\includegraphics[width=0.45\textwidth]{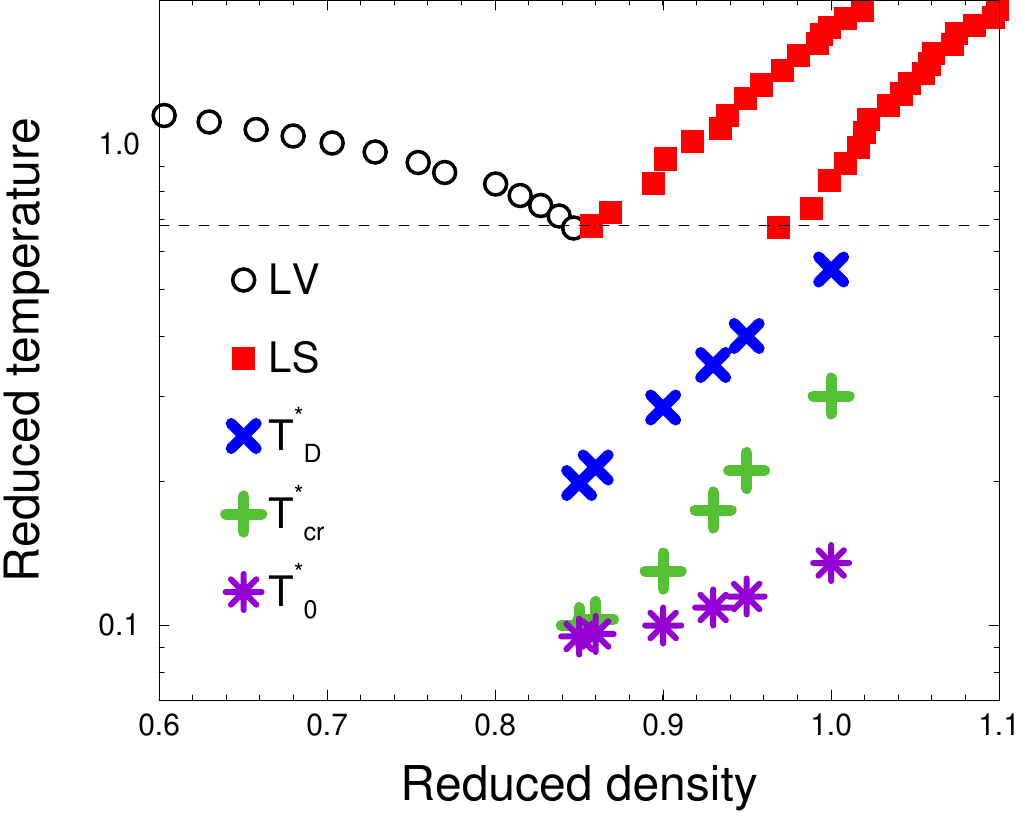}  
\end{tabular}%
\caption{Predicted values of $T^\ast_{\rm D}$ (crosses) $T^\ast_{\rm cr}$ (pluses) and $T^\ast_{\mathrm{0}}$ (stars), together with the liquid branch of liquid-vapor coexistence (circles), and liquid-solid 
coexistence (squares)~\cite{LJwiki}.
The horizontal line indicates the triple-point isotherm. In between crosses and pluses, the supercooled liquid is stable; below $T^\ast_{\rm cr}$, a weak first-order phase transition
makes the ideal glass stable.
Note the logarithmic vertical scale to enhance the visibility of the low-temperature portion.}
\label{fig:PhDia}
\end{figure}

The behavior of $g_{\mathrm{12}}(0)$ for $T^\ast<0.2$ and $\epsilon_{12}=0$ is finally shown as a full line,
with the corresponding $Q$ in the inset. The observed rapid rise of both order parameters, in the absence of any explicit coupling between the replicas, faithfully reflects the behavior expected for an ideal glass trapped in the lowest free energy minimum, with the confinement of different replicas gradually enhanced as the temperature is lowered.
Moreover, the RFOT is characterized by the expected discontinuity 
of the order parameters in the limit of vanishing inter-replica coupling: two different values are reached by $g_{\mathrm{12}}(0)$ for $\epsilon_{12}=\epsilon$ and $\epsilon_{12}=0$. After extinction of $\epsilon_{12}$, we have obtained the dynamical transition temperature $T^\ast_{\rm D}$ by following the fate of $g_{12}(0)$ and $Q$ upon slowly increasing $T^\ast$ between two successive R-HMSA solutions. In doing this, $g_{12}(0)$ and $Q$ progressively decrease, and are found to survive till $T^\ast=0.284$, a threshold to be eventually identified with $T^\ast_{\rm D}$ at $\rho^\ast=0.90$. By repeating the above procedure, we have mapped out the upper boundaries of the ideal glass phase over the density range $0.85 \leq \rho^\ast \leq 1$. 
In Fig.~\ref{fig:PhDia}, our predictions for $T^\ast_{\mathrm{D}}$ (blue crosses) are embodied within the high-density portion of the LJ phase diagram~\cite{LJwiki}. Numerical values are explicitly given in Table~\ref{tab:1}.

\begin{table}[b]
\caption{$T^\ast_{\mathrm{D}}$, $T^\ast_{\mathrm{cr}}$ and $T^\ast_{\mathrm{0}}$ for the densities investigated in this work.}
\label{tab:1}%
\begin{tabular*}{0.40\textwidth}{@{\extracolsep{\fill}}cccc}
\hline\hline
$\rho^\ast$ & $T^\ast_{\rm D}$   & $T^\ast_{\rm cr}$ & $T^\ast_{\mathrm{0}}$\\
\hline
0.85  &  0.198 & 0.100 & 0.095 \\
0.86  &  0.213 & 0.103 & 0.096\\
0.90  &  0.284 & 0.130 & 0.100\\
0.93  &  0.348 & 0.174 & 0.109\\
0.95  &  0.400 & 0.210 & 0.115\\
1.00  &  0.550 & 0.300 & 0.135\\
\hline\hline
\end{tabular*}%
\end{table}

As discussed in Sect.~\ref{subsec:RFOT}, HMSA and R-HMSA numerical predictions, concerning respectively the liquid and the ideal glass phases, a priori ``coexist'' below the dynamical transition temperature. 
The search for the most stable phase requires determining
the free energies within each phase as functions of temperature along some isochores. 

\begin{figure}[t!]
\begin{tabular}{c}
\includegraphics[width=0.48\textwidth]{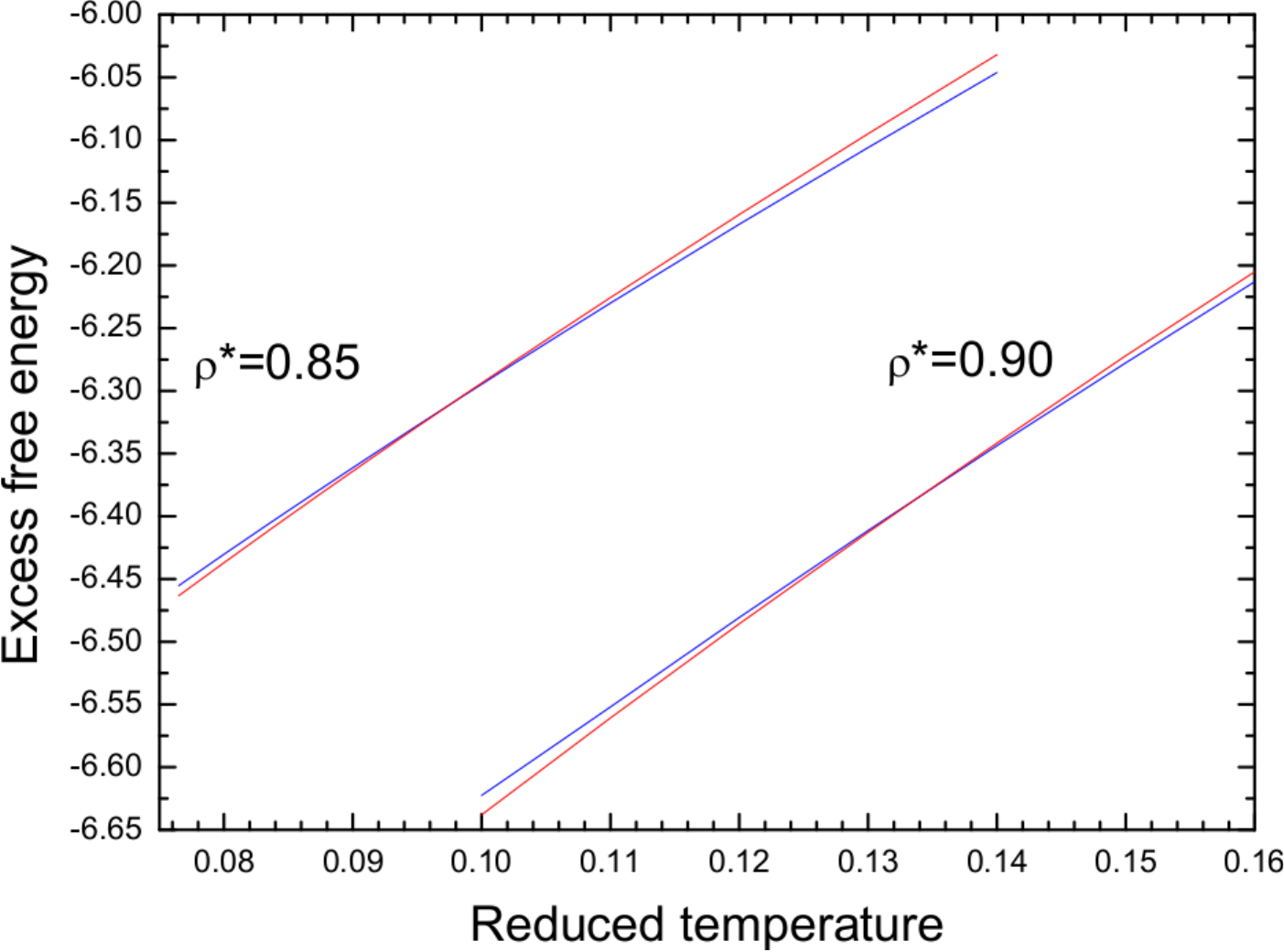}   
\end{tabular}
    \caption{Excess free energies $f_{\rm ex}^{\ast\rm L}$ (blue lines) and $f_{\rm ex}^{\ast\rm G}$ (red lines)
    as functions of $T^\ast$ along the isochores $\rho^\ast=0.85$ and 0.90. Along each isochore, $f_{\rm ex}^{\ast\rm L}$ and $f_{\rm ex}^{\ast\rm G}$ cross at 
    the temperature $T_{\mathrm{cr}}^\ast(\rho^\ast)$.}
\label{fig:Fex}
\end{figure}

Predictions for $f_{\rm ex}^{\ast\rm L}$ and $f_{\rm ex}^{\ast\rm G}$, with generically $f_{\rm ex}^{\ast}=F_{\rm ex}^{\ast}/N\epsilon$~---~%
as deduced accordingly to the thermodynamic integration schemes described Sect.~\ref{subsec:RFOT}~---~are reported along two 
different isochores in Fig.~\ref{fig:Fex}. 
As seen, the free energies corresponding to the liquid and ideal-glass phases are very close to each other, as also observed in a previous HNC study~\cite{BomontPRE2015}.
Along each isochore, $f_{\rm ex}^{\ast\rm L}$ lies below $f_{\rm ex}^{\ast\rm G}$, until a crossing is clearly observed at a temperature $T^\ast_{\mathrm{cr}}$, pointing to a weak first-order transition to the ideal glass, with the latter being stable for $T^\ast< T^\ast_{\mathrm{cr}}$. Our predictions for $T^\ast_{\mathrm{cr}}$ are also shown in Fig.~\ref{fig:PhDia} (green pluses)
along with $T^\ast_{\rm D}$. Numerical values are listed in Table~\ref{tab:1}.

\begin{figure}[t!]
\begin{tabular}{c}
\includegraphics[width=0.48\textwidth]{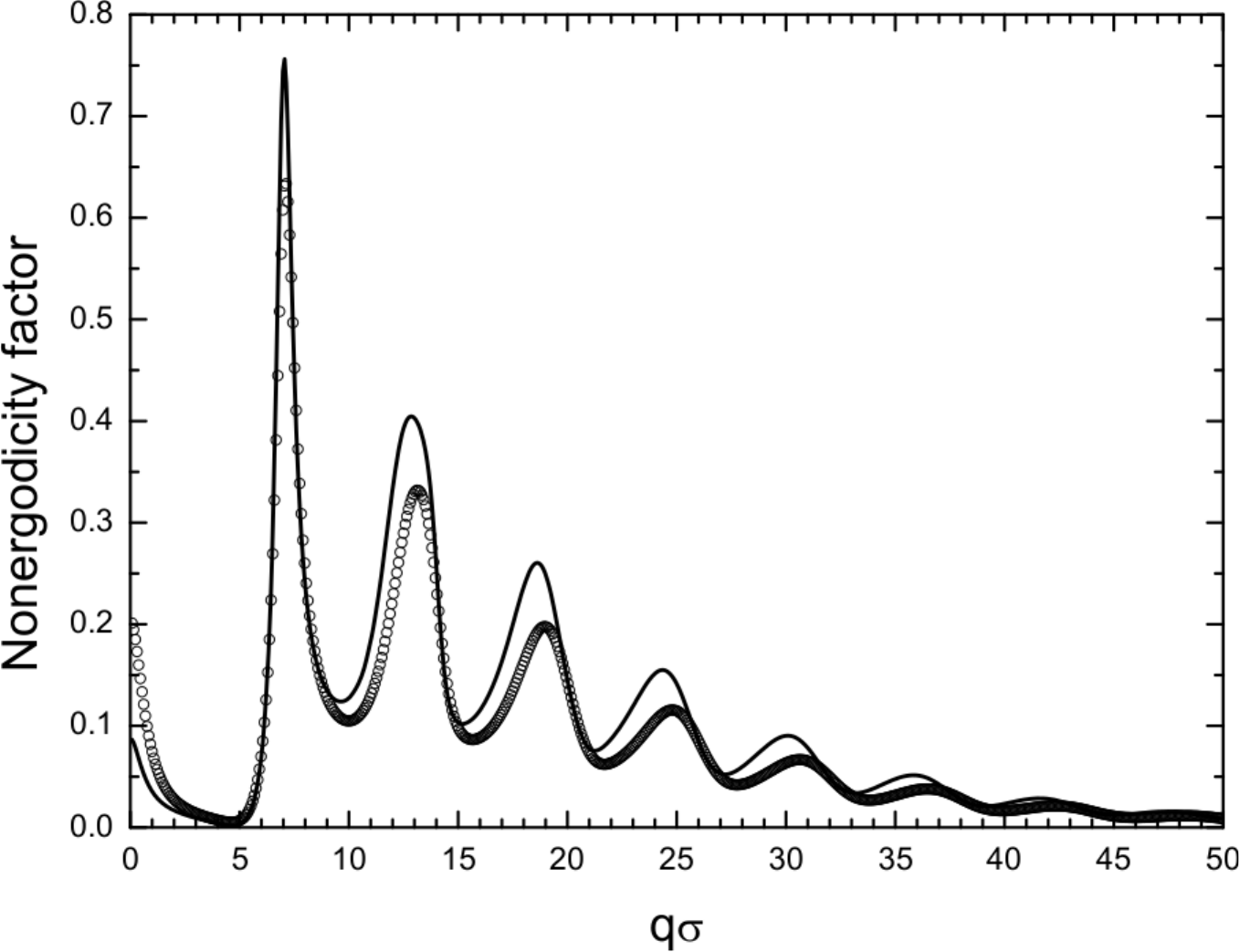} 
  
\end{tabular}
    \caption{Nonergodicity factor $f_{\rm q}$ versus $q\sigma$ calculated by R-HNC (circles) for $\rho^\ast=1$ at its dynamical transition temperature $T^\ast=0.46$ compared to the one obtained by R-HMSA (full line) at the same state point.}
\label{fig:Comp_Fq}
\end{figure}

Within the MCT framework, the glass transition is interpreted as a dynamical transition at a critical temperature from an ergodic (liquid) to a nonergodic phase (glass). The order parameter is the so-called nonergodicity factor $f_{\rm q}$ (the central object of MCT) defined in the reciprocal $q$-space as the long-time limit of the coherent normalized scattering function $F(q,t)$. The fact that $F(q,t\rightarrow \infty)$ does not vanish indicates that the system is in a glass phase. Whatever the potential model, it has been shown that the MCT $f_{\rm q}$ parameter is related to the counterpart of $g_{12}(r)$ in the $q$-space, namely $S_{12}(q)$, so that $f_{\rm q}=S_{12}(q)/S_{11}(q)$, where $S_{11}(q)$ is the intra-replica structure factor~\cite{REVMOD2010}. 

In Fig.~\ref{fig:Comp_Fq}, we show the comparison between $f_{\rm q}$ obtained from replicated HNC (R-HNC) closure with the one from R-HMSA, at the state point $\rho^\ast=1$ and $T^\ast=0.46$, corresponding to the dynamical transition temperature of the former closure relation. As expected, the qualitative behavior of the curves closely resembles that obtained for hard-sphere systems using the R-HNC closure~\cite{REVMOD2010}, which is itself qualitatively consistent with results from MCT~\cite{Gotze1999} and experiments~\cite{Megen1991}. However, compared to R-HNC result, R-HMSA predicts a higher amplitude ($\approx$20$\%$) of the first peak of $f_{\rm q}$, indicating that the quantitative discrepancy reported in Ref.~\citenum{REVMOD2010} can be remedied by employing a more accurate liquid-state closure.

The implementation of the HMSA scheme, together with the ODS split of the pair potential, 
turns to be crucial to penetrate a fundamentally low-temperature regime. Therein, the lowest values $T^\ast_{\rm min}$ attained within the liquid phase (i.e., by HMSA) vary between 0.007 and 0.01, and within the glass phase (i.e., by R-HMSA) between 0.0625 and 0.094, depending on the density. 

As far as a comparison with other theoretical schemes is concerned, we have explored the liquid phase also by using the HMSA coupled with the WCA scheme, to find that the lowest temperature attainable along each isochore is actually much higher than that reached by ODS. For example, by WCA, $T^\ast_{\rm min}(\rho^\ast=0.90) \approx 0.18$, to be compared with the corresponding ODS $T^\ast_{\rm min}=0.009$. In the next Sect.~\ref{sec:4}, we shall discuss in more detail the structural reasons why the HMSA/WCA numerical algorithm fails to converge as a certain low temperature is attained. As for the HNC theory, along $\rho^\ast=0.90$ and 0.95, the lowest temperatures attained are $T^\ast_{\rm min} \approx 0.34$ and $\approx 0.25$, respectively. Below these thresholds, the HNC algorithm fails to converge to a physical solution, returning in particular negative values for the structure factor at zero wavevectors, $S(0)$. In evaluating this failure, one needs to consider that HNC does not require any split of the interaction potential. At first sight, this might appear to be an advantage, insofar as whatever arbitrariness involved in the splitting procedure is avoided. On the other hand, we have shown that the ODS separation of $u(r)$ lets us enforce the zero-separation requirement within the HMSA, and this apparently proves to be essential to improve the theoretical performances. 

Finally, we mention ~---~among various attempts to study theoretically the high-density/low-temperature behavior of LJ~---~ an improved version of the Modified HNC theory~\cite{Rosenfeld}, developed by our late colleague and friend Mimmo Gazzillo~\cite{Gazzillo1993}. In that work, the authors carried out a positive assessment of their structural predictions against simulation data. However, even in that case, the theoretical approach failed to attain a very low temperature regime. For instance, the lowest temperature reached at $\rho^\ast=0.95$ in Ref.~\citenum{Gazzillo1993} was $T^\ast_{\rm min}=0.233$, to be compared with the present  $T^\ast_{\rm min}(\rho^\ast=0.95)=0.007$. 
\begin{figure}[t!]
\begin{tabular}{c}
\includegraphics[width=0.45\textwidth]{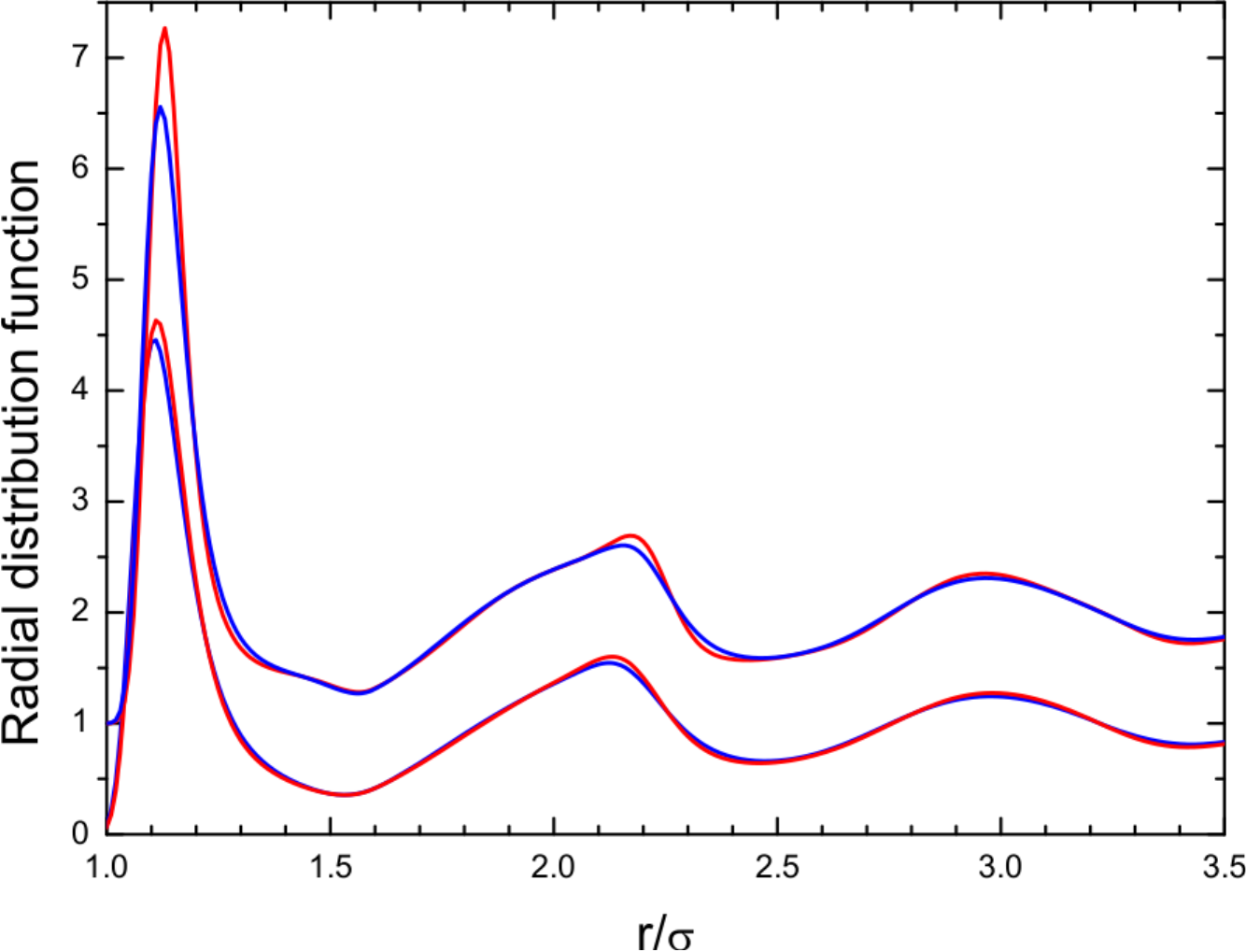}
\end{tabular}
\caption{
Comparison between $g(r)$ in the liquid (blue) and $g_{11}(r)$ in the ideal glass (red) phases for $\rho^\ast=0.9$ 
at $T^\ast=0.2$ (full lines) and $T^\ast=0.1$ (shifted by one along the vertical axis, for clarity)}.
\label{fig:COMP_g}
\end{figure}
As far as simulation studies are concerned, it is well recognized that the investigation of glass-forming systems suffer from the rapidly growing relaxation times near the glass transition, which historically have limited investigations into the regime of very mild supercooling~\cite{Ediger1996}. 
In this context, we recall that the very first attempt to investigate by molecular dynamics the glass state of the LJ fluid, dates back to almost fifty years ago: in Ref.~\citenum{Rahman1976}, Rahman and co-workers were in fact able to generate an amorphous sample by means of a rapid, isochoric quenching from the liquid phase. Since then, a variety of methods~---~as for instance simulated tempering and 
cluster Monte Carlo~\cite{Dress1995}~---~were introduced to deal with issues related to slow equilibration. More recently, 
swap Monte Carlo schemes~---~in which the standard Metropolis sampling is ``augmented'' by swap moves between physically distant particles of different species~---~were shown to substantially speed-up the equilibration in certain glass-forming mixtures, allowing for the study of glassy properties previously out of reach~\cite{swap}. Nevertheless, these huge computational efforts are devoted to the study of polydisperse models only, and
even in this case, the technique is thought to be ``too a primitive approach''~\cite{Berthier2023} requiring substantial improvements to firmly test the validity of the RFOT theory.

\section{Structural properties in the low-temperature regime}\label{sec:4}
\begin{figure}[t!] 
\begin{tabular}{cc}
\includegraphics[width=0.25\textwidth]{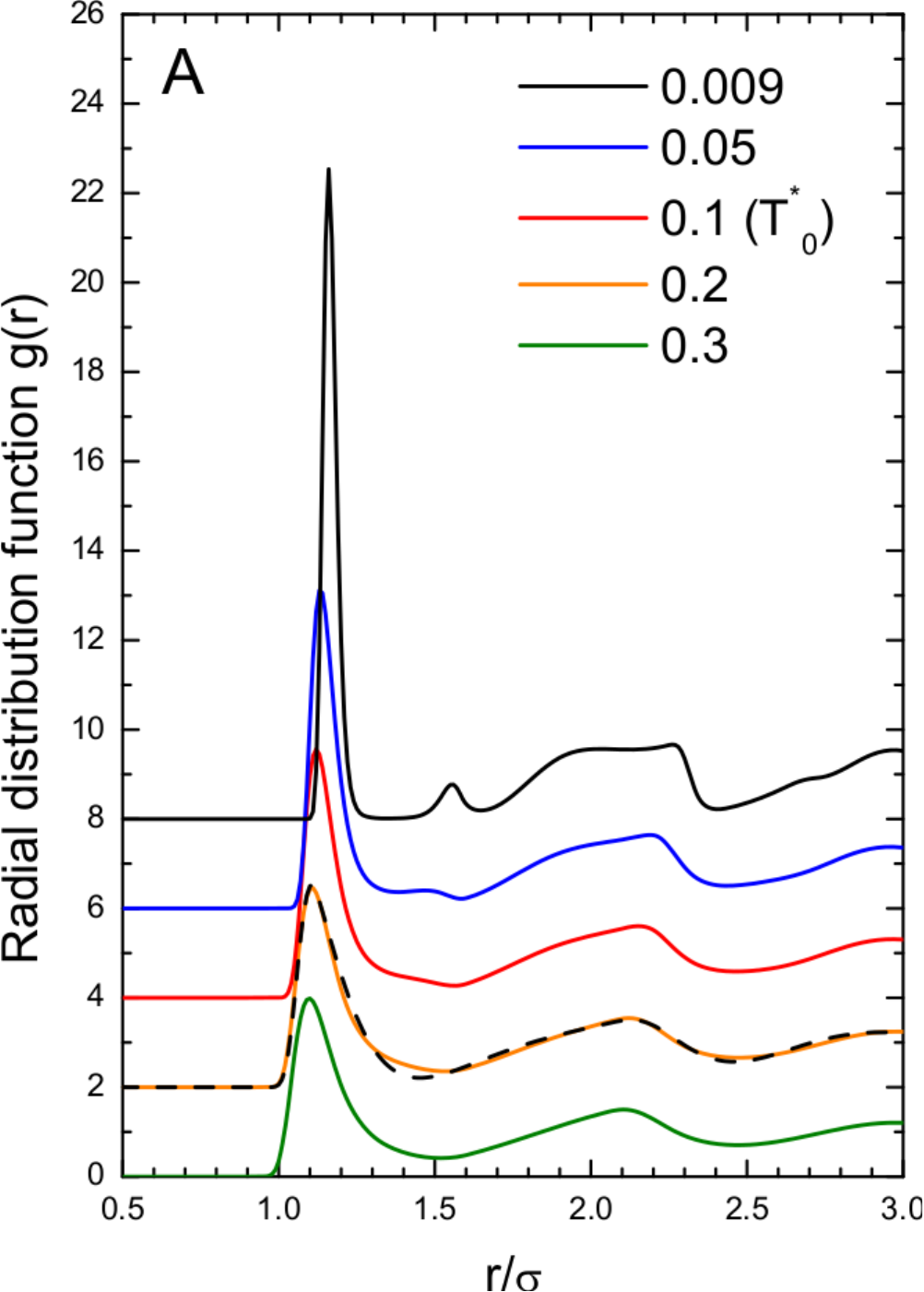} & 
\includegraphics[width=0.23\textwidth]{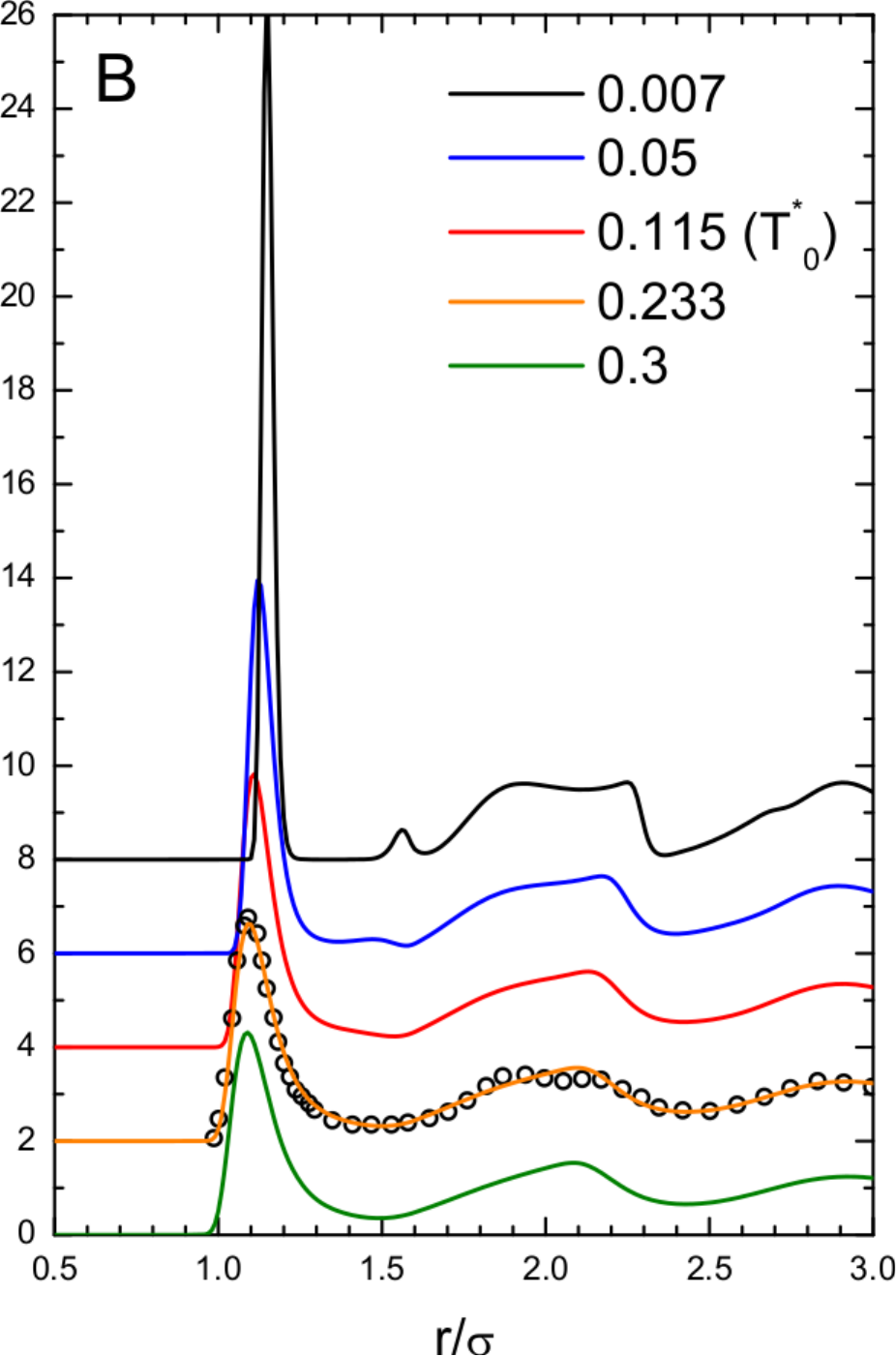} \\
\includegraphics[width=0.25\textwidth]{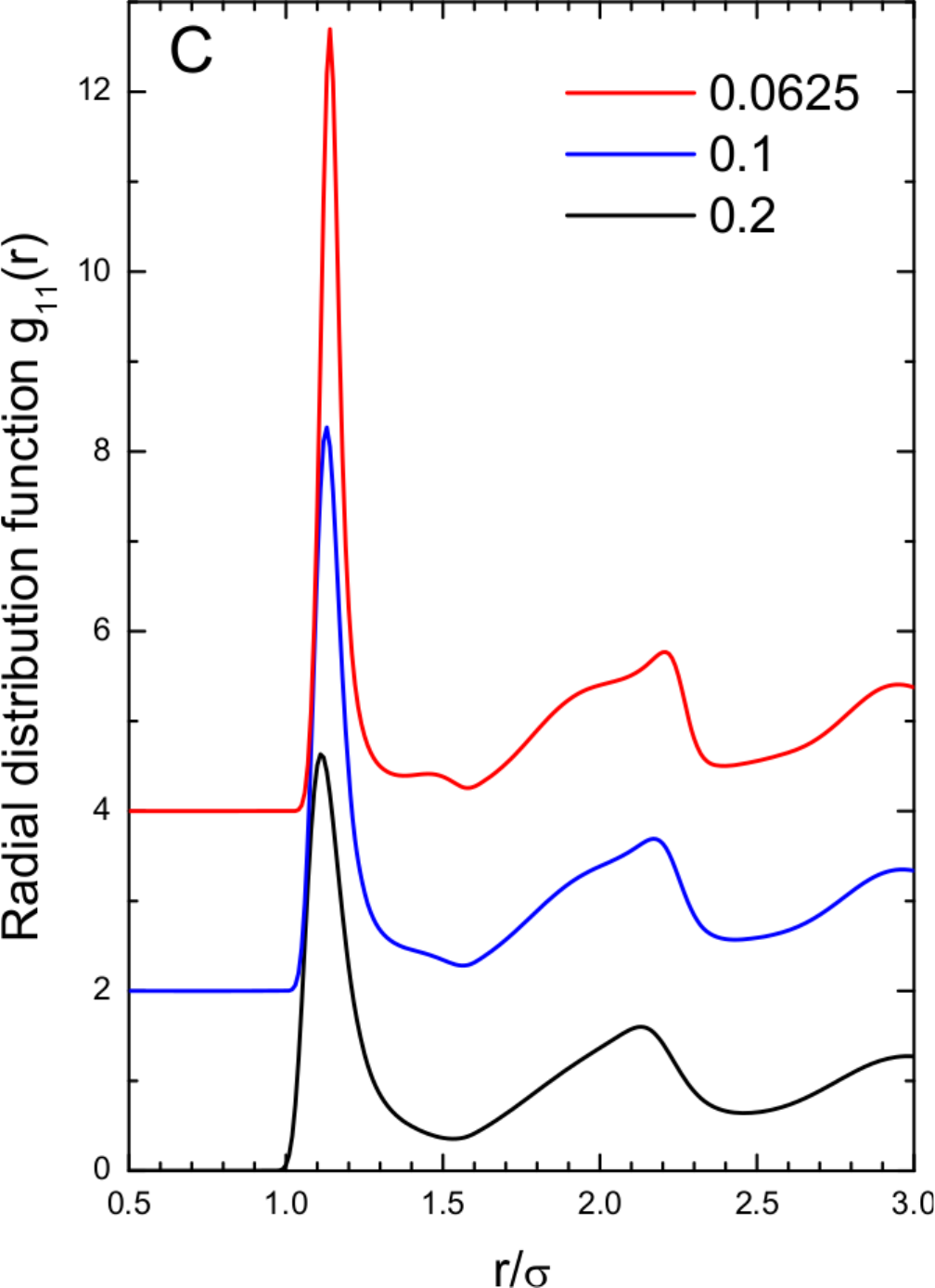} & 
\includegraphics[width=0.228\textwidth]{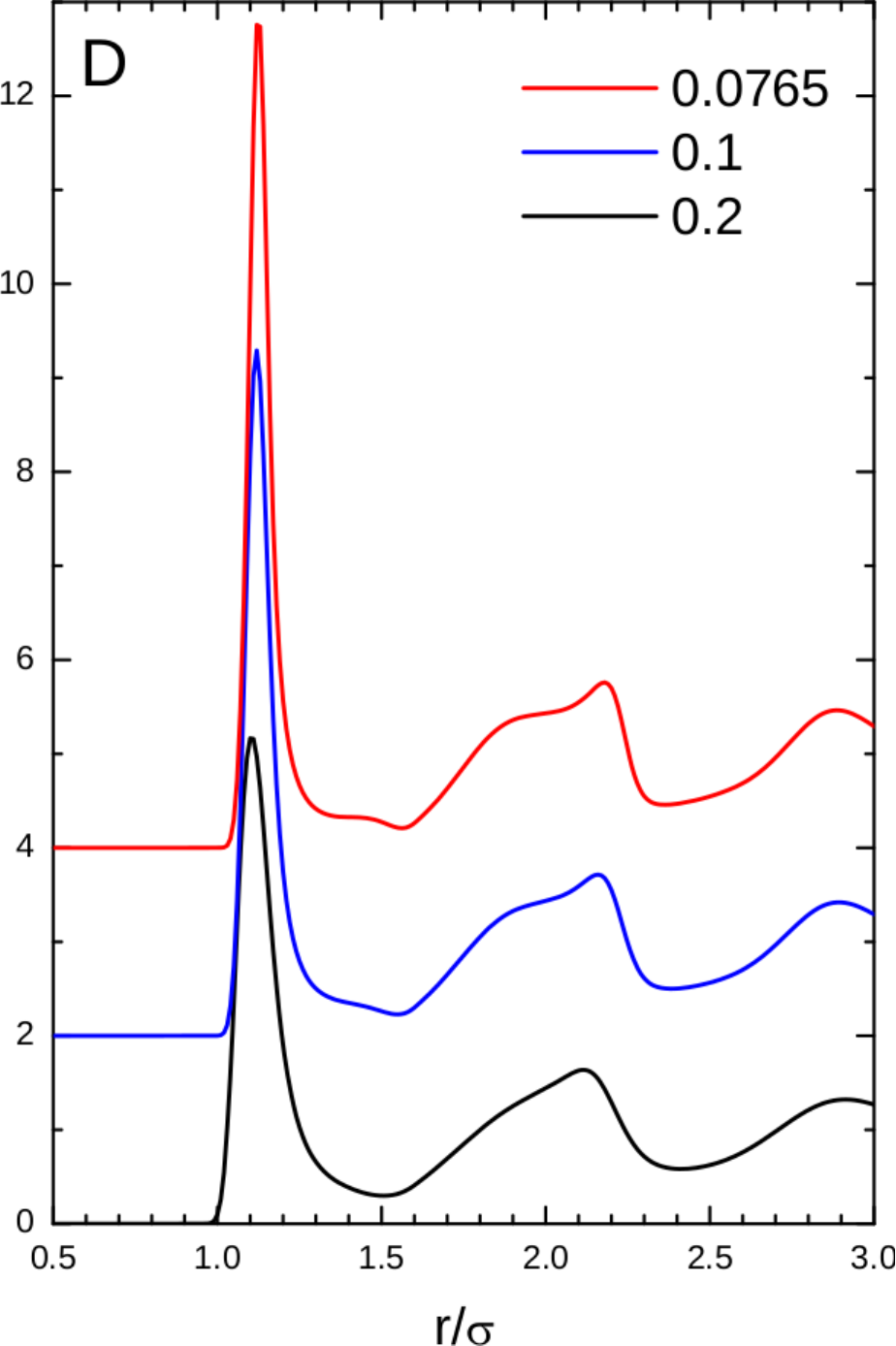} 
\end{tabular}
\caption{Variation of $g(r)$ (A, B) and $g_{11}(r)$ (C, D) across $T_{\mathrm{0}}^\ast$ at $\rho^\ast=0.90$ (A, C) and $\rho^\ast=0.95$ (B, D); temperatures in the legend.
The dashed line in A refers to HMSA predictions coupled with WCA, at $T^\ast=0.2$. Circles in B refer to MD data at $T^\ast=0.233$ of Ref.~\citenum{Gazzillo1993}. 
For clarity, in each panel, all curves but the lowest one are shifted by two along the vertical axis.}
\label{fig:gr}
\end{figure}
We show in this Section that both HMSA and R-HMSA predict the existence of clear-cut structural modifications,
as the fluid proceeds deeply within the supercooled state.
Let us first recall that the thermodynamic integration paths leading to the free energies of Fig.~\ref{fig:Fex}
rely upon the calculation of the internal energy and virial pressure for many different thermodynamic conditions.
Both properties are in turn
obtained from $g(r)$~---~calculated in the liquid phase by HMSA~---~%
and $g_{11}(r)$~---~calculated in the glass phase by R-HMSA.
Therefore, we reasonably expect that the close proximity between $f_{\rm ex}^{\ast\rm L}$ and $f_{\rm ex}^{\ast\rm G}$ 
observed in Fig.~\ref{fig:Fex},
witnesses, below $T^\ast_{\rm D}$ and for
identical thermodynamic conditions, a corresponding similarity  between $g(r)$ and $g_{11}(r)$.
As exemplified in Fig.~\ref{fig:COMP_g}, this is indeed the case. Therein, predictions concern $\rho^\ast=0.90$ and two different
temperatures $T^\ast=0.2$ and $0.1$, falling below $T^\ast_{\rm D}(\rho^\ast=0.90)=0.284$, and across $T^\ast_{\rm cr}(\rho^\ast=0.90)=0.13$. We see
that the main difference between $g(r)$ and $g_{11}(r)$ remains in a tiny discrepancy about the height of the main peak, 
both functions being hardly distinguishable otherwise.

Once shown the close similarity between the structural predictions within the supercooled and the ideal-glass phases,
we proceed to a detailed  analysis in parallel, so as to systematically identify the common features, as the temperature changes.
In Fig.~\ref{fig:gr} we show  $g(r)$ (top) and $g_{11}(r)$ (bottom) for densities $\rho^\ast=0.90$ (left) and $0.95$ (right). 

At $T^\ast=0.3$, $g(r)$ is liquid-like, as $g_{11}(r)$ at $T^\ast=0.2$ does, whatever the density, with two smooth peaks corresponding to the first and second coordination shells. 
As the temperature decreases, the first peak of both $g(r)$ and $g_{11}(r)$ 
becomes more pronounced and narrower, while the second peak becomes progressively asymmetric and broader. 
No appreciable crystalline structure at short-range can be hinted at, yet. 
Further decreasing the temperature produces a remarkable enhancement of structure. 
In fact, at a peculiar temperature $T^\ast_{\mathrm{0}}(\rho^\ast)$ (see Fig.~\ref{fig:PhDia} and Table I), both $g(r)$ and $g_{11}(r)$ exhibit a change of slope on the left-hand side of the first minimum, in such a way that its shape turns from convex to concave. 
As a prominent feature, this change heralds, below $T^\ast_{\mathrm{0}}$, the formation of a
well-resolved additional peak in between the original first two peaks, then slightly moving towards larger distances upon further cooling. Such a peak is known to signal the development of a ``distorted crystal''~\cite{Rabani1999}.  
When this occurs, the short-range structure is sufficiently compact to appear solid-like in character, without any propensity for a long-range order to show up at the same time. We note that the position of such an additional peak is not exactly pertinent to a perfect fcc arrangement~---~and there is no reason for such an expectation~---~%
thereby indicating that the short-range order is ``quasi-fcc''. We recall in this instance that, based on the Ackland-Jones analysis~\cite{Ackland2006}, as applied to disordered systems, Abraham has shown~\cite{Abraham2015} that in the glass region the local packing environment around an individual atom becomes more crystalline-like and that this locally-crystalline structure becomes highly interconnected. 

\begin{figure}[t!]
\begin{tabular}{c}
\includegraphics[width=0.45\textwidth]{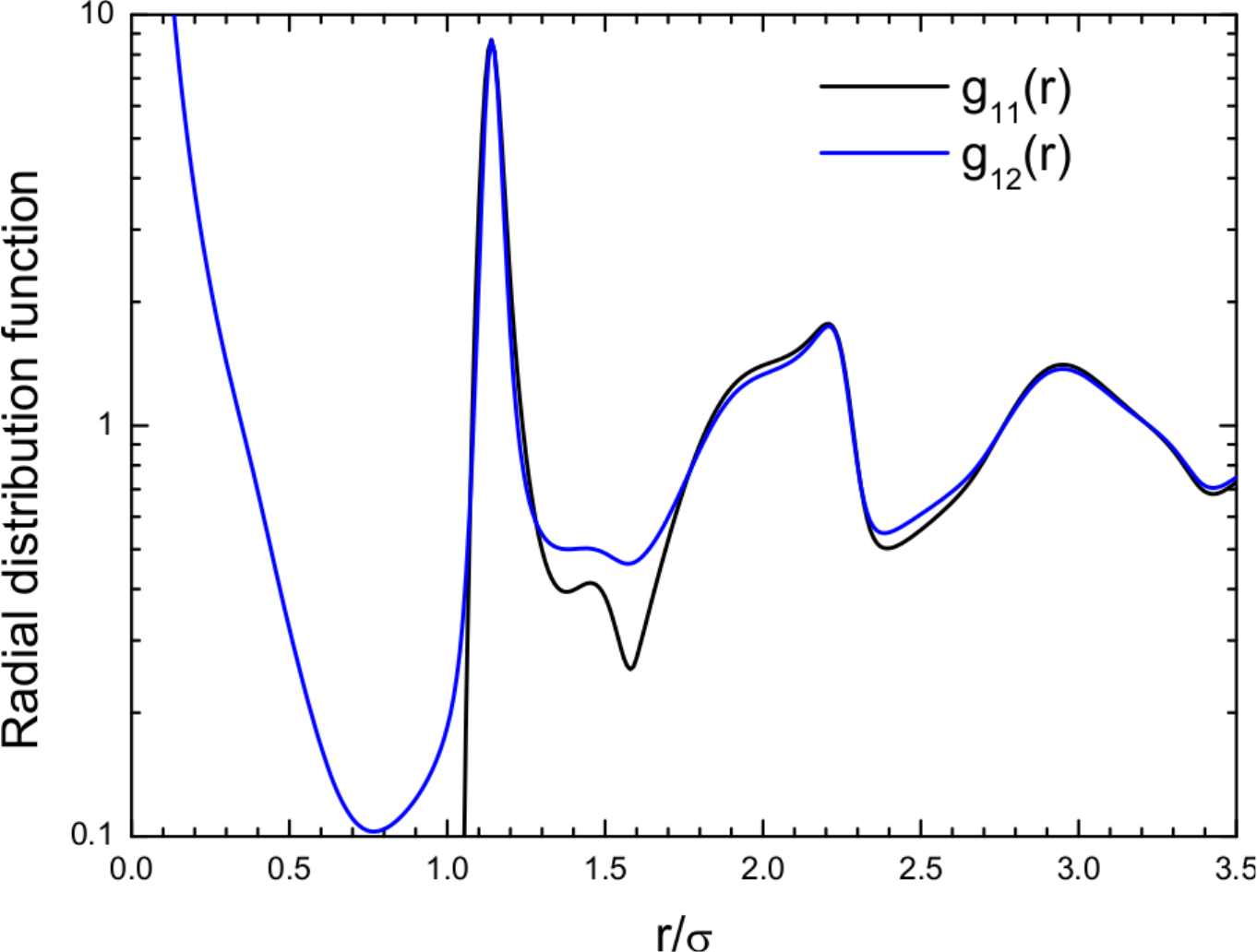}
\end{tabular}
\caption{
Radial distribution functions $g_{11}(r)$ (black) and $g_{12}(r)$ (blue) in the ideal-glass phase for $\rho^\ast=0.9$ at $T^\ast=0.0625$. While $g_{12}(r=0)=1531$ (not shown), both of these functions oscillate in phase at larger distances and exhibit the additional peak at about $r/\sigma=1.45$, between the original first two.}
\label{fig:gr12}
\end{figure}

In Fig.~\ref{fig:gr12} are shown $g_{11}(r)$ and $g_{12}(r)$ for $\rho^\ast=0.90$ at $T^\ast=0.0625$. As seen, both functions oscillate in phase, according to the expectation that a unique underlying disordered equilibrium configuration of the ideal-glass exists, around which the atoms of the two replicas vibrate. Moreover, $g_{12}(r)$ signals as well the local structural change occurring within the ideal-glass, with the presence of the additional peak at about $r/\sigma=1.45$, in between the original first two.

To close this discussion, 
 HMSA/WCA $g(r)$ at $\rho^\ast=0.90$ and $T^\ast=0.2$ is shown as a dashed line in Fig.~\ref{fig:gr}A. As seen, it compares well with ODS $g(r)$, but in the proximity of the first minimum. This discrepancy will progressively transform, as the temperature is lowered, into a negative 
 ``correlation hole'', eventually triggering the failure of the HMSA/WCA numerical integration algorithm. Finally, 
the study in Ref.~\citenum{Gazzillo1993} offers us the opportunity to positively assess our theoretical predictions for $g(r)$ against molecular dynamics results, at least in one case. This comparison is shown at $\rho^\ast=0.95$ in Fig.~\ref{fig:gr}B, for $T^\ast=0.233$, i.e. the lowest temperature attained in that paper. Looking at small discrepancies in the figure, we observe that the deformation of the second peak, barely arising in the MD $g(r)$, turns out to be simply delayed to lower temperatures by HMSA.
We stress once again that~---~to the best of our knowledge~---~no much more simulation structural data are available 
in the low-temperature regime to compare with, for the monodisperse Lennard-Jones model.

\begin{figure}[t]
\begin{tabular}{c}

\includegraphics[width=0.48\textwidth]{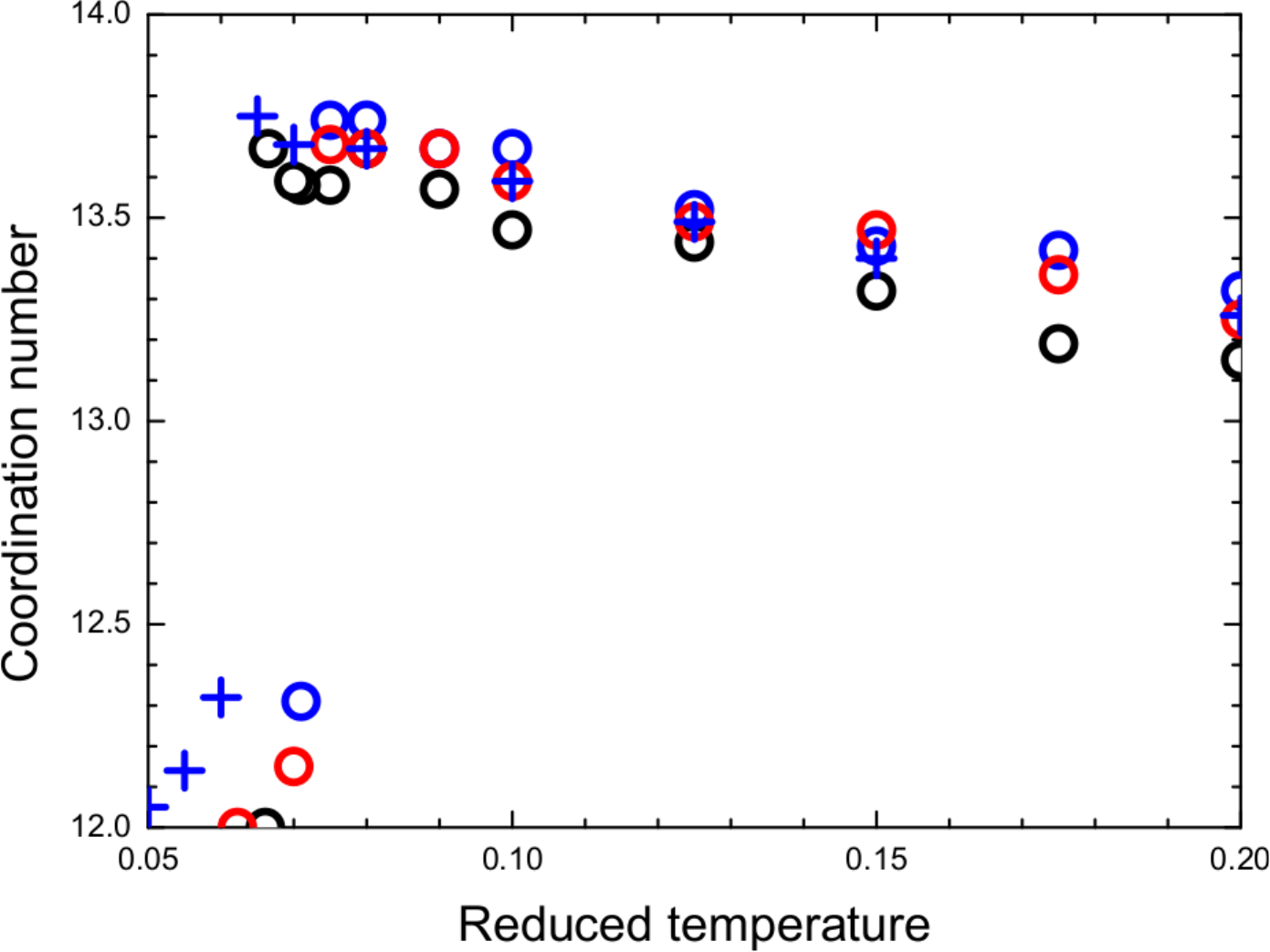}
\end{tabular}
\caption{Coordination number $\langle n \rangle$ vs $T^\ast$ within the liquid phase at $\rho^\ast=0.90$ (crosses) and within the ideal-glass phase along the isochores $\rho^\ast=0.86$ (black circles), $\rho^\ast=0.90$ (red circles) and $\rho^\ast=0.93$ (blue circles).}
\label{fig:nr}
\end{figure}

We now re-interpret in Fig.~\ref{fig:nr} the structural information contained in Fig.~\ref{fig:gr}, in terms of the coordination number $\langle n \rangle$, as calculated by integrating both $\rho^\ast g(r)$ and $\rho^\ast g_{11}(r)$ from 0 to their first mimimum. 
The coordination number describes the local structure of an atom's immediate surroundings (i.e. the first coordination shell), as consisting of a certain number of other atoms, that form a cage around. 
Upon cooling, the latter can be modified by vibrational motion.
Above $T^\ast_{\mathrm{0}}$, the width of the first shell is broad enough, to practically exclude any atomic occupancy reminiscent of a crystalline packing at short distances. In other words, the first coordination shell must be somehow distorted~\cite{Charbonneau2014}, to account for the fact that thirteen or more atoms can be accommodated, instead of twelve, marking a perfect fcc order. However~---~if we recall what just said apropos Fig.~\ref{fig:gr}~---~as the temperature decreases, the main peak shrinks, and this process is then accompanied by the development of an additional peak. This witnesses how, for $T^\ast<T^\ast_{\mathrm{0}}$, the first-neighbor cage reorganizes in such a way to expel atoms in excess with respect to the expectation number typical of a fcc order.
In fact, since $\rho^\ast g(r)$ is integrated till the position of its very first minimum, the appearance of the additional peak is thus accompanied by a drastic decrease of $\langle n \rangle$. 
 Figure~\ref{fig:nr} quantitatively illustrates the changes occurring in $\langle n \rangle$ at $\rho^\ast=0.90$ in the liquid phase and at $\rho^\ast=0.86$, $0.90$ and $0.93$ in the ideal-glass phase along each isochore with decreasing temperature: the abrupt loss of neighbors in excess below $T_0^\ast(\rho)$ is clearly visible. At the lowest temperatures reached here, $\langle n \rangle$ turns exactly 12. To close, upon cooling, $\langle n \rangle$ undergoes a discontinuous drop-down to a much lower value, at variance with the continuous change observed in $g(r)$.

\section{Conclusions}\label{sec:conclusions}  

We have studied the high-density/low-temperature thermodynamic states of the Lennard-Jones fluid, by means of a two-replica version of the HMSA theory, combined with the ODS split of the pair potential. First, we have shown that our approach successfully reaches a very low-temperature regime. Down there, we have first identified the locus $T_{\rm D}^\ast(\rho)$, representing the upper boundary for the putative existence of 
the ideal-glass phase, and then, by free-energy calculations, the locus  $T_{\rm cr}^\ast(\rho)$~---~with $T_{\rm cr}^\ast < T_{\rm D}^\ast$~---~below which the ideal glass 
becomes effectively stable over the supercooled liquid (see Fig.~\ref{fig:PhDia}). The transition between these two states sets as a weak first-order phase transition.

Looking at the radial distribution function, we have further identified a density-dependent temperature $T_0^\ast(\rho)$, 
across which structural modifications indicate the onset of a solid-like short-range order, without any long-range order. 
In the absence of better observations based on modern computer simulations, we conjecture that such a distorted-solid structure could be 
that pertinent to the ideal glass at very low temperature.
Ultimately, our predictions witness how structural correlations are by no means ``mute'' and/or ``uninteresting spectators''~\cite{Guiselin2022}, insofar as profound structural changes accompany the system along the path towards lower and lower temperatures, in close correspondence with the thermodynamic behavior. 

We like to consider our study as shedding some fresh light on the behavior of perhaps the most extensively studied prototype model for liquids, turning a century now~\cite{LJ1924}, and since then the subject of relentless interest. 
In the future, we envisage the possibility to extend our formalism to binary mixtures, so to study other standard glass-formers, such as the Kob-Andersen model mentioned in the Introduction.

\acknowledgments 
The authors wish to thank Jean-Pierre Hansen for fruitful discussion and for a careful reading of the manuscript during all stages of preparation.


\section*{References}
\begin{thebibliography}{99}

\bibitem{Zanotto2018} 
E. D. Zanotto and D. R. Cassar, J. Chem. Phys. \textbf{149}, 024503 (2018).

\bibitem{Zanotto2023} 
R. S. Welch, E. D. Zanotto, C. J. Wilkinson, D. R. Cassar, M.~Montazerian, and J. C. Mauro, Acta Materialia \textbf{254}, 118994 (2023).  

\bibitem{Kob1998} 
W. Kob, Annu. Rev. of Comput. Phys. III, pp. 1-43 (1995).

\bibitem{Guiselin2022} 
B. Guiselin, G. Tarjus, and L. Berthier, Phys. Chem. Glasses \textbf{63}, 136 (2022).

\bibitem{Biroli2013} 
G. Biroli and J. P. Garrahan, J. Chem. Phys. \textbf{138}, 12A301 (2013).

\bibitem{Angell1995} 
C. A. Angell, Science \textbf{267}, 1924 (1995).

\bibitem{Debenedetti2001} 
P. G. Debenedetti and F. H. Stillinger, Nature (London) \textbf{410}, 259 (2001).

\bibitem{BerthierBiroli2011} 
L. Berthier and G. Biroli, Rev. Mod. Phys. \textbf{83}, 587 (2011).

\bibitem{Gotze2009} 
W. G\"otze, ``Complex Dynamics of Glass-Forming Liquids: A Mode-Coupling Theory'' (Oxford University Press, New York, 2009).

\bibitem{Royall} 
L. Ortlieb, T. S. Ingebrigtsen, J. E. Hallett, F. Turci, and C.~P.~Royall, Nat. Commun. \textbf{14}, 2621 (2023).

\bibitem{Kob1994} 
W. Kob and H. C. Andersen, Phys. Rev. Lett. \textbf{73}, 1376 (1994). 

\bibitem{Kob1995A} 
W. Kob and H. C. Andersen, Phys. Rev. E \textbf{51}, 4626 (1995).

\bibitem{Kob1995B} 
W. Kob and H. C. Andersen, Phys. Rev. E \textbf{53}, 4134 (1995).

\bibitem{Kauzmann1948} 
W. Kauzmann, Chem. Rev. \textbf{43}, 219 (1948).

\bibitem{Stillinger} 
F. H. Stillinger, J. Chem. Phys. \textbf{88}, 7818 (1988).

\bibitem{Tanaka2003} 
H. Tanaka, Phys. Rev. E \textbf{68}, 011505 (2003).



\bibitem{Martin2025} 
A. Martin and M. Thuo, Angew. Chem. \textbf{137}, e202423536 (2025).

\bibitem{Royall2018}
C. P. Royall, F. Turci, S. Tatsumi, J. Russo, and J. Robinson, J. Phys.: Condens. Matter \textbf{30}, 363001 (2018).

\bibitem{Wolynes}
For a detailed discussion of RFOT, see the review articles collected in P.~G. Wolynes and V. Lubchenko (Editors), ``Structural Glasses and Supercooled Liquids: Theory, Experiment, and Applications'' (John Wiley, Hoboken, 2012).

\bibitem{Franz1995}
S. Franz and G. Parisi, Journal de Physique I \textbf{5}, 1401 (1995).

\bibitem{Mezard1996}
M. M\'{e}zard, Journal de Physique I \textbf{5}, 1401 (1995).

\bibitem{Franz1997}
S. Franz and G. Parisi, Phys. Rev. Lett. \textbf{79}, 2486 (1997).

\bibitem{Franz1998}
S. Franz and G. Parisi, Physica A \textbf{261}, 317 (1998).

\bibitem{Cardenas1998}
M. Cardenas et al J. Phys. A: Math. Gen. \textbf{31} L163 (1998). 

\bibitem{Cardenas1999}
M. Cardenas et al J. Chem. Phys. \textbf{110}, 1726–1734 (1999). 

\bibitem{Coluzzi1999}
B. Coluzzi et al  J. Chem. Phys. \textbf{111}, 9039–9052 (1999). 


\bibitem{Hansen2013}
J.-P.~Hansen and I.~R.~McDonald, ``Theory of Simple Liquids'', 4th edition (Academic Press, Oxford, 2013). 

\bibitem{BomontACP2008}
J.-M. Bomont, Adv. Chem. Phys. \textbf{139}, 1 (2008). DOI: 10.1002/9780470259498.ch1.

\bibitem{Janssen2024}
I. Pihlajamaa and L. M. C. Janssen, Phys. Rev. E \textbf{110}, 044608 (2024).

\bibitem{Berthier2022}
B. Guiselin, L. Berthier, and G. Tarjus, SciPost Phys. \textbf{12}, 091 (2022).

\bibitem{BomontEPL2014}
J.-M. Bomont, G. Pastore, and J.-P. Hansen, Europhys. Lett. \textbf{105}, 36003 (2014).

\bibitem{BomontJCP2014}
J.-M. Bomont, J.-P. Hansen, and G. Pastore, J. Chem. Phys. \textbf{141}, 174505 (2014).

\bibitem{BomontMOLPHYS2015}
J.-M. Bomont and G. Pastore, Molec. Phys. \textbf{113}, 2770 (2015).
 
\bibitem{BomontPRE2015}
J.-M. Bomont, J.-P. Hansen, and G. Pastore, Phys. Rev. E \textbf{92}, 042316 (2015).

\bibitem{Bomont2017}
J.-M. Bomont, J.-P. Hansen, and G. Pastore, ``Advances in the Computational Sciences`` \textbf{7}, 108 (2017). 
DOI: 10.1142/9789813209428\_0007

\bibitem{BomontJCP2017}
J.-M. Bomont, J.-P. Hansen, and G. Pastore, J. Chem. Phys. \textbf{146}, 114504 (2017).
 
\bibitem{BomontJCP2019}
J.-M. Bomont, J.-P. Hansen, and G. Pastore, J. Chem. Phys. \textbf{150}, 154504 (2019).
  
\bibitem{BomontPRE2022}
J.-M. Bomont, C. N. Likos, and J.-P. Hansen, Phys. Rev. E. \textbf{105}, 024607 (2022). 

\bibitem{BomontJCP2024}
J.-M. Bomont, G. Pastore, D. Costa, G. Malescio, G. Munaò, and S. Prestipino, J. Chem. Phys. \textbf{160}, 214504 (2024).

\bibitem{Franz2013}
S. Franz, H. Jaquin, G. Parisi, P. Urbani and F. Zamponi, J. Chem. Phys. \textbf{138}, 12A540 (2013).

\bibitem{Zerah} 
G. Z\'{e}rah and J.-P. Hansen, J. Chem. Phys. \textbf{84}, 2336 (1986).

\bibitem{Bomont2001} 
J.-M. Bomont and J.-L. Bretonnet, J. Chem. Phys. \textbf{114}, 4141 (2001).

\bibitem{Roland2024} 
R. Kjellander,``Statistical Mechanics of Liquids and Solutions – Intermolecular Forces, Structure and Surface Interactions'', 1st edition (CRC Press - Taylor and Francis, 2024). DOI: 10.1201/9781003286882 

\bibitem{Bomont1997} 
J.-M. Bomont, N. Jakse, and J. L. Bretonnet, J. Chem. Phys. \textbf{107}, 8030 (1997).

\bibitem{Bomont1998}
J.-M. Bomont, N. Jakse, and J. L. Bretonnet, Phys. Rev. B \textbf{57}, 10217 (1998).

\bibitem{Mezard2012} 
M. M\'{e}zard and G. Parisi, ``Glasses and Replicas'', Chap.4, pp. 151-191 (John Wiley and Sons, New-York, 2012).

\bibitem{Lee1996} 
L. L. Lee, D. Ghonasgi, and E. Lomba, J. Chem. Phys. \textbf{104}, 8058 (1996).

\bibitem{DH1}
D.-M.~Duh and A.~D.~J.~Haymet, J. Chem. Phys. \textbf{97}, 7716 (1992).

\bibitem{DH2}
D.-M.~Duh and A.~D.~J.~Haymet, J. Chem. Phys. \textbf{103}, 2625 (1994).

\bibitem{Restrepo1}
M. Llano-Restrepo and W. G. Chapman, J. Chem. Phys. \textbf{97}, 2046 (1992).

\bibitem{Restrepo2}
M. Llano-Restrepo and W. G. Chapman, J. Chem. Phys. \textbf{100}, 5139 (1994).

\bibitem{Vompe1}
G. A. Martynov and A. G. Vompe, Phys. Rev. E \textbf{47}, 1012 (1993).

\bibitem{Vompe2}
A. G. Vompe and G. A. Martynov, J. Chem. Phys. \textbf{100}, 5249 (1994).

\bibitem{WCA} 
J. D. Weeks, D. Chandler and H. C. Andersen, J. Chem. Phys. \textbf{54}, 4931 (1970).

\bibitem{Bomont_2003_JCP_1} 
J.-M. Bomont and J.-L. Bretonnet, J. Chem. Phys. \textbf{119}, 2188 (2003).

\bibitem{Bomont_2003_molphys} 
J.-M. Bomont and J.-L. Bretonnet, Molec. Phys. \textbf{101}, 3249 (2003).

\bibitem{Bomont_2003_JCP_2}
J.-M. Bomont, J. Chem. Phys. \textbf{119}, 11484 (2003).

\bibitem{Bomont_2004_JCP}
J.-M. Bomont and J.-L. Bretonnet, J. Chem. Phys. \textbf{121}, 1548 (2004).

\bibitem{Bomont_2006_JCP}
J.-M. Bomont, J. Chem. Phys. \textbf{124}, 206101 (2006).

\bibitem{Bomont_2007_JCP}
J.-M. Bomont and J.-L. Bretonnet, J. Chem. Phys. \textbf{126}, 214504 (2007).

\bibitem{Lee_c_2010} 
L. L. Lee, M. C. Hara, S. J. Simon, F. S. Ramos, A. J. Winkle, and J.-M. Bomont, J. Chem. Phys. \textbf{132}, 074505 (2010).


\bibitem{Gillan1979} 
M. J. Gillan, Molec. Phys. \textbf{38}, 1781 (1979).

\bibitem{Charbonneau2014}
P. Charbonneau, J. Kurchan, G. Parisi, P. Urbani, and F. Zamponi, Nat. Comm. \textbf{5}, 3725 (2014).

\bibitem{Bretonnet2016}
J.-L. Bretonnet, Molec. Phys. \textbf{114}, 2868 (2016).

\bibitem{Mezard1987}
M. M\'{e}zard, G. Parisi, and M. A. Virasoro, Spin Glass Theory and Beyond (World Scientific, Singapore, 1987).

\bibitem{Monasson1995}
R. Monasson, Phys. Rev. Lett. \textbf{75}, 2847 (1995).

\bibitem{Costa2002} 
D. Costa, G. Pellicane, M. C. Abramo, and C. Caccamo, J. Chem. Phys. \textbf{118}, 304 (2002).

\bibitem{BerthierPRE2015} 
L. Berthier and R. L. Jack Phys. Rev. Lett.\textbf{ 114}, 205701 (2015).

\bibitem{LJwiki}
\url{https://en.wikipedia.org/wiki/Lennard-Jones_potential}

\bibitem{REVMOD2010} 
G. Parisi and F. Zamponi, Rev. Mod. Phys. \textbf{82}, 789 (2010).

\bibitem{Gotze1999}
W. G\"otze, J. Phys.: Condens. Matter \textbf{11}, A1 (1999).

\bibitem{Megen1991}
W. van Megen, S. M. Underwood and P. N. Pusey, Phys. Rev. Lett. \textbf{67}, 1586 (1991).

\bibitem{Rosenfeld} 
Y. Rosenfeld and N. W. Ashcroft, Phys. Rev. A \textbf{20}, 1208 (1979).
 
\bibitem{Gazzillo1993}
D. Gazzillo and R. G. Della Valle, J. Chem. Phys. \textbf{99}, 6915 (1993).

\bibitem{Ediger1996} 
M. D. Ediger, C. A. Angell, and S. R. Nagel, J. Phys. Chem. \textbf{100}, 13200 (1996).

\bibitem{Rahman1976}
A. Rahman,  M. J. Mandell, and J. P. McTague, J. Chem. Phys. \textbf{64}, 1564 (1976).
\bibitem{Dress1995} 
C. Dress and W. Krauth, J. Phys. A \textbf{28}, L587–L601 (1995).
\bibitem{swap}
L. Berthier, E. Flenner, C. J. Fullerton, C. Scalliet and M. Singh, J. Stat. Mech. Theory Exp. \textbf{2019}, 064004 (2019).
\bibitem{Berthier2023}
L. Berthier and D. R. Reichman, Nature Reviews Physics \textbf{5}, 102 (2023).






\bibitem{Rabani1999}
E. Rabani, J. D. Gezelter, and B. J. Berne, Phys. Rev. Lett.  \textbf{82}, 3649 (1999).

\bibitem{Ackland2006}
G. Ackland and A. Jones, Phys. Rev. B \textbf{73}, 054104 (2006).

\bibitem{Abraham2015}
F. F. Abraham, \url{https://arxiv.org/abs/1504.05751} (2015).

\bibitem{LJ1924} 
J. E. Jones, Proc. R. Soc. Lond. A \textbf{106}, 441 (1924).












\end{thebibliography}
\end{document}